\documentclass[
aip,
amsmath,
amssymb,
reprint]{revtex4-1}

\usepackage{graphicx}
\usepackage{dcolumn}
\usepackage{bm}
\usepackage{epstopdf}
\usepackage{svg}

\usepackage[utf8]{inputenc}
\usepackage[T1]{fontenc}
\usepackage{mathptmx}

\begin{document}

\preprint{AIP/123-QED}

\title[Controlling acoustic waves using magneto-elastic Fano resonances]{Controlling acoustic waves using magneto-elastic Fano resonances}

\author{O. S. Latcham}
    \affiliation{University of Exeter, Stocker Road, Exeter, EX4 4QL, United Kingdom}
\author{Y. I. Gusieva}
    \affiliation{Igor Sikorsky Kyiv Polytechnic Institute, 37 Prosp. Peremohy, Kyiv, 03056, Ukraine}
\author{A. V. Shytov}
    \affiliation{University of Exeter, Stocker Road, Exeter, EX4 4QL, United Kingdom}
\author{O. Y. Gorobets}
    \affiliation{Igor Sikorsky Kyiv Polytechnic Institute, 37 Prosp. Peremohy, Kyiv, 03056, Ukraine}
\author{V. V. Kruglyak}
    \email{V.V.Kruglyak@exeter.ac.uk}
    \affiliation{University of Exeter, Stocker Road, Exeter, EX4 4QL, United Kingdom}
\date{\today}

\begin{abstract}
We propose and analyze theoretically a class of energy-efficient magneto-elastic devices for analogue signal processing.  The signals are carried by transverse acoustic waves while the bias magnetic field controls their scattering from a magneto-elastic slab.  By tuning the bias field, one can alter the resonant frequency at which the propagating acoustic waves hybridize with the magnetic modes, and thereby control transmission and reflection coefficients of the acoustic waves. The scattering coefficients exhibit Breit-Wigner/Fano resonant behaviour akin to inelastic scattering in atomic and nuclear physics.  Employing oblique incidence geometry, one can effectively enhance the strength of magneto-elastic coupling, and thus countermand the magnetic losses due to the Gilbert damping.  We apply our theory to discuss potential benefits and issues in realistic systems and suggest routes to enhance performance of the proposed devices.
\end{abstract}


\maketitle 

Optical and, more generally, wave-based computing paradigms gain momentum on a promise to replace and complement the traditional semiconductor-based technology.\cite{Feitelson_1988} The energy savings inherent to non-volatile memory devices has spurred the rapid growth of research in magnonics, \cite{Kruglyak_2010, Nikitov_2015} in which spin waves\cite{Akhiezer_1968} are exploited as a signal or data carrier. Yet, the progress is hampered by the magnetic loss (damping).\cite{Krivoruchko_2015, Azzawi_2017} Indeed, the propagation distance of spin waves is rather short in ferromagnetic metals while low-damping magnetic insulators are more difficult to structure into nanoscale devices.  In contrast, the propagation distance of acoustic waves is typically much longer than that of spin waves at the same frequencies.\cite{Collins_1984} Hence, their use as the signal or data carrier could reduce the propagation loss to a tolerable level.  Notably, one could control the acoustic waves using a magnetic field by coupling them to spin waves within magnetostrictive materials.\cite{Kittel_1958, Bommel_1959, Dreher_2012} To minimize the magnetic loss, the size of such magneto-acoustic functional elements should be kept minimal.  This implies coupling propagating acoustic waves to confined spin wave modes of finite-sized magnetic elements. As we show below this design idea opens a route towards hybrid devices combining functional benefits of magnonics\cite{Kruglyak_2010, Nikitov_2015} with the energy efficiency of phononics.\cite{Collins_1984, Li_2012, Maldovan_2013}

The phenomena resulting from interaction between coherent spin and acoustic waves have already been addressed in the research literature: the spin wave excitation of propagating acoustic waves\cite{Collins_1984, Hollander_2018, Bauer_2018, Thingstad_2019} and vice versa,\cite{Kittel_1958, Li_2017, Gowtham_2015, Ulrichs_2017} acoustic parametric pumping of spin waves,\cite{Gurevich_1965, Keshtgar_2014, Chowdhury_2015} magnon-phonon coupling in cavities\cite{Litvinenko_2015, Zhango_2016, Kong_2019} and mode locking,\cite{Wang_1970} magnonic-phononic crystals,\cite{Nikitov_2012, Graczyk_2017} Bragg scattering of spin waves from a surface acoustic wave induced grating,\cite{Chumak_2010, Kryshtal_2017_1, Kryshtal_2017_2} topological properties of magneto-elastic excitations,\cite{Thingstad_2019, Takahashi_2016} acoustically driven spin pumping and spin Seebeck effect,\cite{Uchida_2011, Polzikova_2018} and optical excitation and detection of magneto-acoustic waves.\cite{Yahagi_2014, Kats_2016, Berk_2017, Yang_2018, Deb_2018, Mondal_2018, Hashimoto_2018} However, studies of the interaction between propagating acoustic waves and spin wave modes of finite-sized magnetic elements, which are the most promising for applications, have been relatively scarce to date.\cite{Dreher_2012, Yahagi_2014, Berk_2017, Mondal_2018}
\begin{figure}[ht!]
    \centering
    \includegraphics[width=85mm]{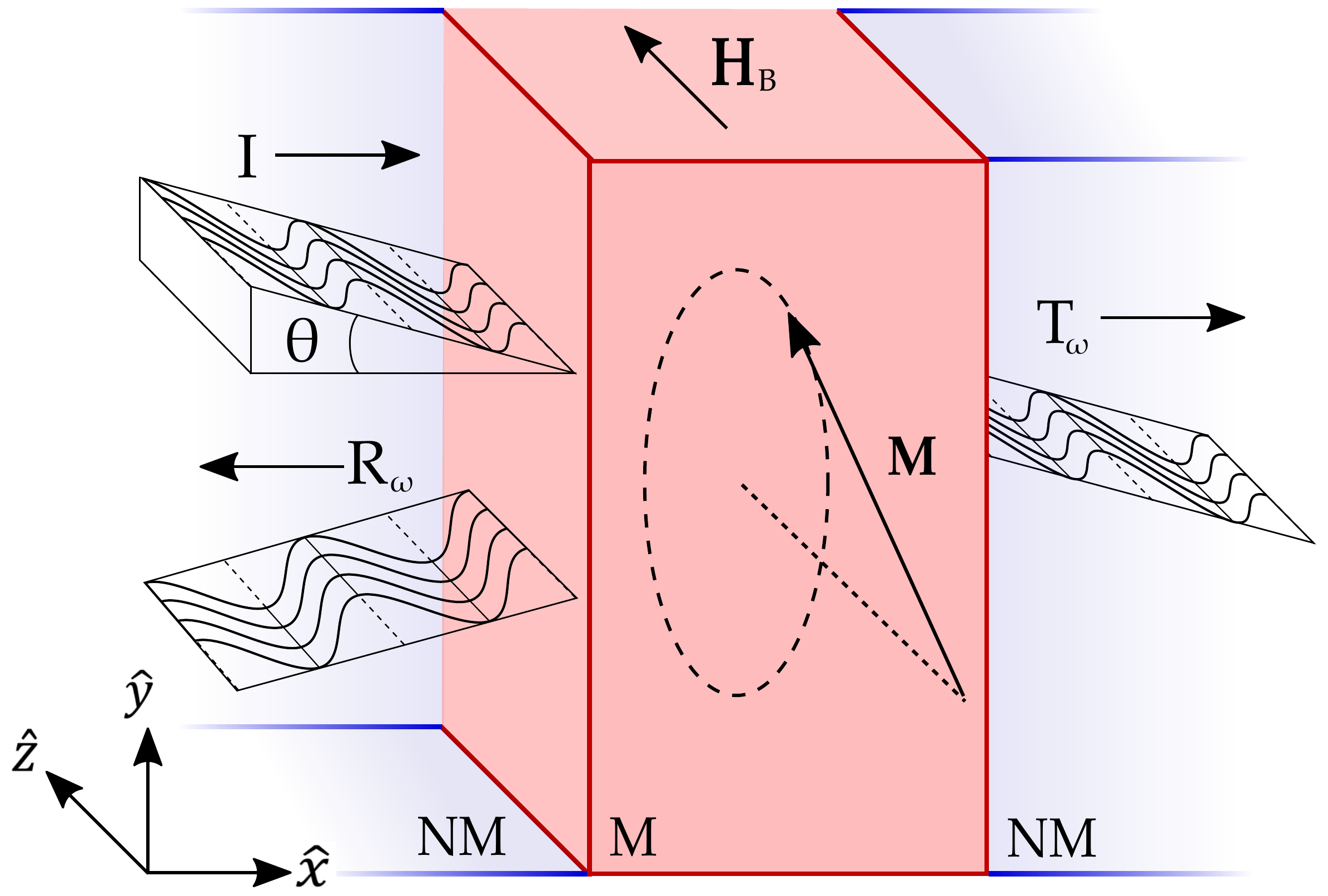}
    \caption{\label{fig:fig1}The prototypical magneto-elastic resonator
    is a thin magnetic slab (M) of width $\delta$, biased by an external field~${\bf H}_{\mathrm{B}}$, and embedded into a non-magnetic (NM) matrix. The acoustic wave with amplitude $I$ incident at angle $\theta$ induces precession of the magnetisation vector~${\bf M}$ via the magneto-elastic coupling. As a result, the wave is partly transmitted and reflected, with respective amplitudes~$T_\omega$ and~$R_\omega$.}
\end{figure}

Here, we explore theoretically the class of magneto-acoustic devices in which the signal is carried by acoustic waves while the magnetic field controls its propagation via the magnetoelastic interaction in thin isolated magnetic inclusions as shown in Fig.~\ref{fig:fig1}. By changing the applied magnetic field, one can alter the frequency at which the incident acoustic waves hybridize with the magnetic modes of the inclusions.  Thereby, one can control the acoustic waves by the resonant behaviour of Breit-Wigner and Fano resonances in the magnetic inclusion.\cite{Limonov_2017} We find that the strength of the resonances is suppressed by the ubiquitous magnetic damping in realistic materials, but this can be mitigated by employing oblique incidence geometry. To compare magneto-acoustic materials for such devices, we introduce a figure of merit.  The magneto-elastic Fano resonance is identified as most promising in terms of frequency and field tuneability. To enhance resonant behaviour, we explore the oblique incidence as a means by which to enhance the figure of merit.

We consider the simplest geometry in which magneto-elastic coupling can affect sound propagation. A ferromagnetic slab ("magnetic inclusion") of thickness $\delta$, of the order of 10 nm, is embedded within a non-magnetic medium (Fig.\ref{fig:fig1}). The slab is infinite in the $Y-Z$ plane, has saturation magnetization $M_{\mathrm{s}}$, and is biased by the applied field $\mathbf{H}_\mathrm{B} = H_{\mathrm{B}}\hat{\bf z}$. Due to the magneto-elastic coupling, this equilibrium configuration is perturbed by shear stresses in the  $xz$- and $yz$ planes associated with the incident acoustic wave. 

To derive the equations of motion, we represent the magnetic energy density $F$ of the magnetic material as a sum of the magneto-elastic $F_{\mathrm{ME}}$ and purely magnetic $F_{\mathrm{M}}$ contributions.\cite{Comstock_1963} Taking into account the Zeeman and demagnetizing energies, we write $F_{\mathrm{M}} = -\mu_{0}\bm{H}_\mathrm{B}\bm{M} + \frac{\mu_{0}}{2}(N_{x}M_{x}^{2}+N_{y}M_{y}^{2})$, where $N_{x(y)}$ are the demagnetising coefficients, $N_{x}+N_{y}=1$, $\bm{M}$ is the magnetization and $\mu_{0}$ is the magnetic permeability. In a crystal of cubic symmetry, the magnetoelastic contribution takes the form\cite{Kamra_2015}
\begin{equation}
    \label{eqn:eqn1}
    F_{\mathrm{ME}} = \frac{B}{M_{\mathrm{s}}^{2}} \sum_{i\neq j}M_{i}M_{j}u_{ij} + \frac{B'}{M_{\mathrm{s}}^{2}} \sum_{i} M_{i}^{2} u_{ii}, \;\;\;\;\; i, j = x, y, z ,
\end{equation}
\noindent
where $B'$ and $B$ are the linear isotropic and anisotropic magneto-elastic coupling constants, respectively.\cite{Callen_1965} The strain tensor is $u_{jk} = \frac{1}{2}\left(\partial_{j}U_{k} + \partial_{k}U_{j}\right)$, where $U_{j}$ are the displacement vector components.
To maximize the effect of the
coupling $B$, 
we consider a transverse acoustic plane wave incident on the slab from the left and polarized
along the bias field, so that $U_x = U_y = 0$,  $U_z = U(x, y, t)$.
The 
non-vanishing components of the strain tensor
are $u_{xz} = \frac{1}{2}\partial_{x} U$
and $u_{yz} = \frac{1}{2}\partial_{y}U$, 
and~$F_{\mathrm{ME}}$ is linear in both~${\bm M}$ and~$U$:
\begin{equation}
    \label{eqn:eqn2}
    F_{\mathrm{ME}} = \frac{B}{M_{\mathrm{s}}}(M_{x}u_{xz}+M_{y}u_{yz}).
\end{equation}

The magnetization dynamics in the slab is due to the effective magnetic field, $\mu_{0}\bm{H}_{\mathrm{eff}} = -\delta F/\delta \bm{M}$. We define $\bm{m}$ as the small perturbation of the magnetic order, i.e. $|\bm{m}| \ll M_{\mathrm{s}}$. Linearizing the Landau-Lifshitz-Gilbert equation,\cite{Akhiezer_1968} we write
\begin{eqnarray}
    \label{eqn:eqn3}
    -\frac{\partial m_{x}}{\partial t} &&= \gamma \mu_{0}(H_{\mathrm{B}}+N_{y}M_{\mathrm{s}})m_{y} +\gamma B \frac{\partial U}{\partial y}+ \alpha\frac{\partial m_{y}}{\partial t},\\
    \label{eqn:eqn4}
    \frac{\partial m_{y}}{\partial t} &&= \gamma \mu_{0} \left(H_{\mathrm{B}}+N_{x}M_{\mathrm{s}}\right)m_{x}+\gamma B \frac{\partial U}{\partial x} + \alpha\frac{\partial m_{x}}{\partial t},
\end{eqnarray}
\noindent
where $\gamma$ is the gyromagnetic ratio and $\alpha$ is the Gilbert damping constant.
To describe the acoustic wave, we include the magneto-elastic contribution to the stress, $\sigma^{\mathrm{(ME)}}_{jk} = \delta F_{ME}/\delta u_{jk}$,
 into the momentum balance equation:
\begin{equation}
    \label{eqn:eqn5}
    \rho \frac{\partial^{2}U}{\partial t^2}=\frac{\partial}{\partial x}\left(C\frac{\partial U}{\partial x}+\frac{B}{M_{\mathrm{s}}}m_{x}\right)+\frac{\partial}{\partial y}\left(C\frac{\partial U}{\partial y} + \frac{B}{M_{\mathrm{s}}}m_{y}\right),
\end{equation}
\noindent
where $C = c_{44}$ is the shear modulus and $\rho$ is the mass density. The non-magnetic medium is described by Eq.(\ref{eqn:eqn5}) with~$B=0$.

Since the values of $C$, $B$, and~$N_{x,y}$  are constant within each individual material, we shall seek solutions of the equations in the form of plane waves $U, m_{x(y)} \propto \mathrm{exp}[i(k_{\omega, x}x+k_{\omega, y}y-\omega t)]$. From herein, we consider all variables in the Fourier domain. For the magnetization precession in the magnetic layer driven by the acoustic wave, we thus obtain
\begin{eqnarray}
    \label{eqn:eqn6}
  m_{x} &=& \frac{\gamma B \left(\omega k_{\omega, y} + i \widetilde\omega_{y}k_{\omega, x}\right)}{\omega^{2} - \widetilde\omega_{x}\widetilde\omega_{y}}U,
 \\
    \label{eqn:eqn7}
    m_{y} &=& \frac{i\gamma B \left(\widetilde\omega_{x}k_{\omega, y} + i \omega k_{\omega, x}\right)}{\omega^{2} - \widetilde\omega_{x}\widetilde\omega_{y}}U,
\end{eqnarray}
\noindent
where we have denoted $\omega_{x(y)} = \gamma\mu_{0}(H_{\mathrm{B}}+N_{x(y)}M_{\mathrm{s}})$ and $\widetilde\omega_{x(y)}=\omega_{x(y)} - i\omega\alpha$. The complex-valued wave number $\bm{k}_{\omega, x}$ is given by the dispersion relation
\begin{equation}
    \label{eqn:eqn8}
    k_{\omega, x}^{2} = \frac{\frac{\rho}{C}\omega^2 \left(\omega^2 - \widetilde\omega_{x}\widetilde\omega_{y}\right) - k_{\omega, y}^{2}\left(\omega^2 - \widetilde\omega_{x}\widetilde\omega_{y} + \frac{\gamma B^{2}}{M_{\mathrm{s}}C}\widetilde\omega_{x}\right)}{\left[\omega^2 - \widetilde\omega_{x}\widetilde\omega_{y} + \frac{\gamma B^{2}}{M_{\mathrm{s}}C}\widetilde\omega_{y}\right]},
\end{equation}
\noindent
 where $k_{\omega, y}$ is equal to that of the incident wave, and the branch with $\mathop{\mathrm{Im}}\nolimits k_{\omega, x} > 0$ describes a forward wave decaying into the slab. Eq.~(\ref{eqn:eqn8}) describes the hybridization between acoustic waves and magnetic precession at frequencies close to ferromagnetic resonance (FMR) at frequency $\omega_{\mathrm{FMR}}$, with linewidth $\Gamma_{\mathrm{FMR}}$. The frequency at which the precession amplitudes (Eqs.~(\ref{eqn:eqn6}) and~(\ref{eqn:eqn7})) diverge is given by the condition $\left(\omega_{\mathrm{FMR}} + i\Gamma_{\mathrm{FMR}}/2\right)^{2} = \widetilde\omega_{x}\widetilde\omega_{y}$. In the limit of small $\alpha$, this yields $\omega_{\mathrm{FMR}}=\omega_{x}\omega_{y}$ and $\Gamma_{\mathrm{FMR}} = \alpha(\omega_{x}+\omega_{y})$. Away from the resonance, Eq.~(\ref{eqn:eqn8}) gives the linear dispersion of acoustic waves. 
In the non-magnetic medium ($B=0$),
one finds
$k_{0}^{2} = \omega^2\rho_{0}/C_{0}$. Here and below, the subscript '0' is used to mark quantities pertaining to the non-magnetic matrix. 

To calculate the reflection and transmission coefficients,
$R_{\omega}$ and $T_{\omega}$, 
for a magnetic inclusion, we introduce the mechanical impedance as $Z=i\sigma_{xz}/\omega U_{\omega}$. 
Solution of the wave matching problem
can then be expressed via the ratio of load ($Z_{\mathrm{ME}}$) and source ($Z_{0}$) impedances. For impedances in the forward (F) and backward (B) directions in the magnetic slab, we find 
\begin{equation}
\label{eqn:eqn9}
    Z_{\omega, \mathrm{ME}}^\mathrm{(F/B)} =  \frac{Ck_{\omega, x}}{\omega}\left(1+\frac{\gamma B^{2}}{CM_{\mathrm{s}}}\frac{\widetilde\omega_{y}\mp i\omega\frac{k_{\omega, y}}{k_{\omega, x}}}{\omega^{2}-\widetilde\omega_{x}\widetilde\omega_{y}}\right).
\end{equation}
Here, the `-' and `+' signs correspond to (F) and (B), respectively.
For the non-magnetic material, Eq.~(\ref{eqn:eqn9}) recovers the usual acoustic impedance\cite{Brekhovskikh_1997} $Z_{0} = \mathrm{cos}\theta\sqrt{\rho_{0} C_{0}}$. Due to magnon-phonon hybridization, $\mathop{\mathrm{Re}}\nolimits Z^{(F/B)}_{\omega, \mathrm{ME}}$ diverges at $\omega_{\mathrm{FMR}}$ and vanishes at a nearby frequency $\omega_{\mathrm{ME}}$. For $\alpha = 0$, the latter is given by
\begin{equation}
    \label{eqn:omegame}
    \omega_{\mathrm{ME}} =  \sqrt{\omega_{x}\omega_{y} - \frac{\gamma B^{2}}{M_{\mathrm{s}}C}\omega_{y}}.
\end{equation}

Reflection $R_{\omega}$ and transmission $T_{\omega}$ coefficients are then found via the well-known relations\cite{Brekhovskikh_1997} as
\begin{align}
\label{eqn:eqn11}
R_{\omega} &= \frac{(\widetilde{\eta}_{\omega}+1)(1-\eta_{\omega})\sin(k_{\omega, x}\delta)}{(\widetilde{\eta}_{\omega}\eta_{\omega}+1)\sin(k_{\omega, x}\delta) + i(\eta_{\omega} + \widetilde{\eta}_{\omega})\cos(k_{\omega, x}\delta)}, \\
\label{eqn:eqn12}
T_{\omega} &= \frac{i(\eta_{\omega}+\widetilde{\eta}_{\omega})}{(\widetilde{\eta}_{\omega}\eta_{\omega}+1)\sin(k_{\omega, x}\delta) + i(\eta_{\omega} + \widetilde{\eta}_{\omega})\cos(k_{\omega, x}\delta)},
\end{align}
\noindent
where $\delta$ is the thickness of the magnetic inclusion,  $\eta_{\omega} = Z_{\mathrm{ME}}^\mathrm{(F)}/Z_{0}$ and $\widetilde{\eta}_{\omega} = {Z}_{\mathrm{ME}}^\mathrm{(B)}/Z_{0}$.\cite{Born_1964} In close proximity to the resonance, the impedances changes rapidly. Expanding Eq.~(\ref{eqn:eqn11}) near $\omega_{\mathrm{ME}}$ in the limit $k_{\omega}\delta\ll1$, we obtain
\begin{figure*}[ht!]
   \centering
    \includegraphics[width=170mm]{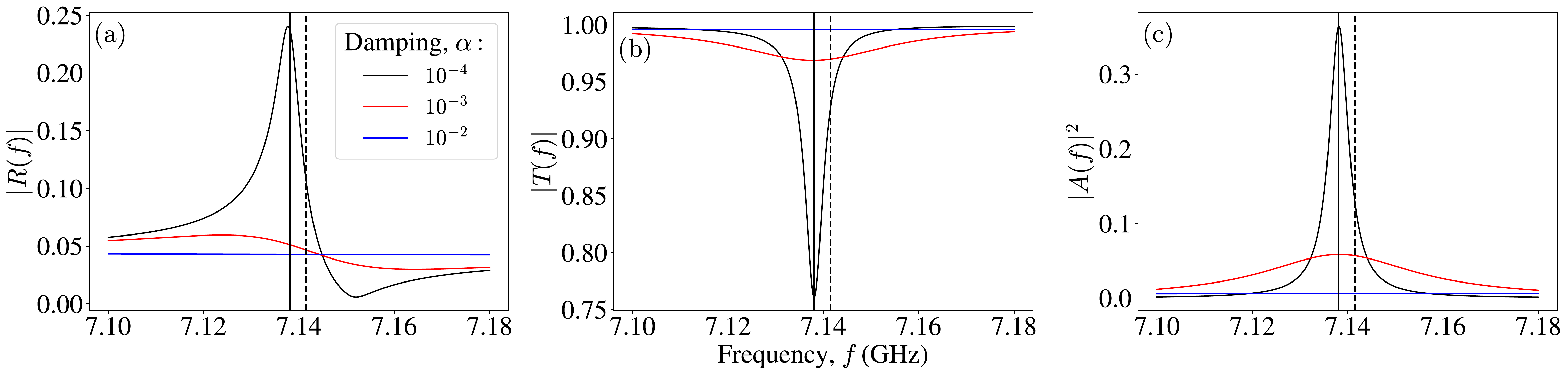}
    \caption{\label{fig:fig2}The frequency dependence of the absolute values of (a) reflection and (b) transmission coefficients and (c) absorbance is shown for a 20nm thick magnetic inclusion.  The vertical dashed and solid black lines represent the ferromagnetic resonance frequency $\omega_{\mathrm{FMR}}$ and magneto-elastic resonance frequency $\omega_{\mathrm{ME}}$ respectively. The non-magnetic and magnetic materials are assumed to be silicon nitride and cobalt, respectively, with parameters given in the text.  The bias field is $\mu_{0}H_{\mathrm{B}}=50$mT, which leads to $f_{\mathrm{ME}}\approx7.138$ GHz.}
\end{figure*}
\begin{figure}[hb!]
    \centering
    \includegraphics[width=85mm]{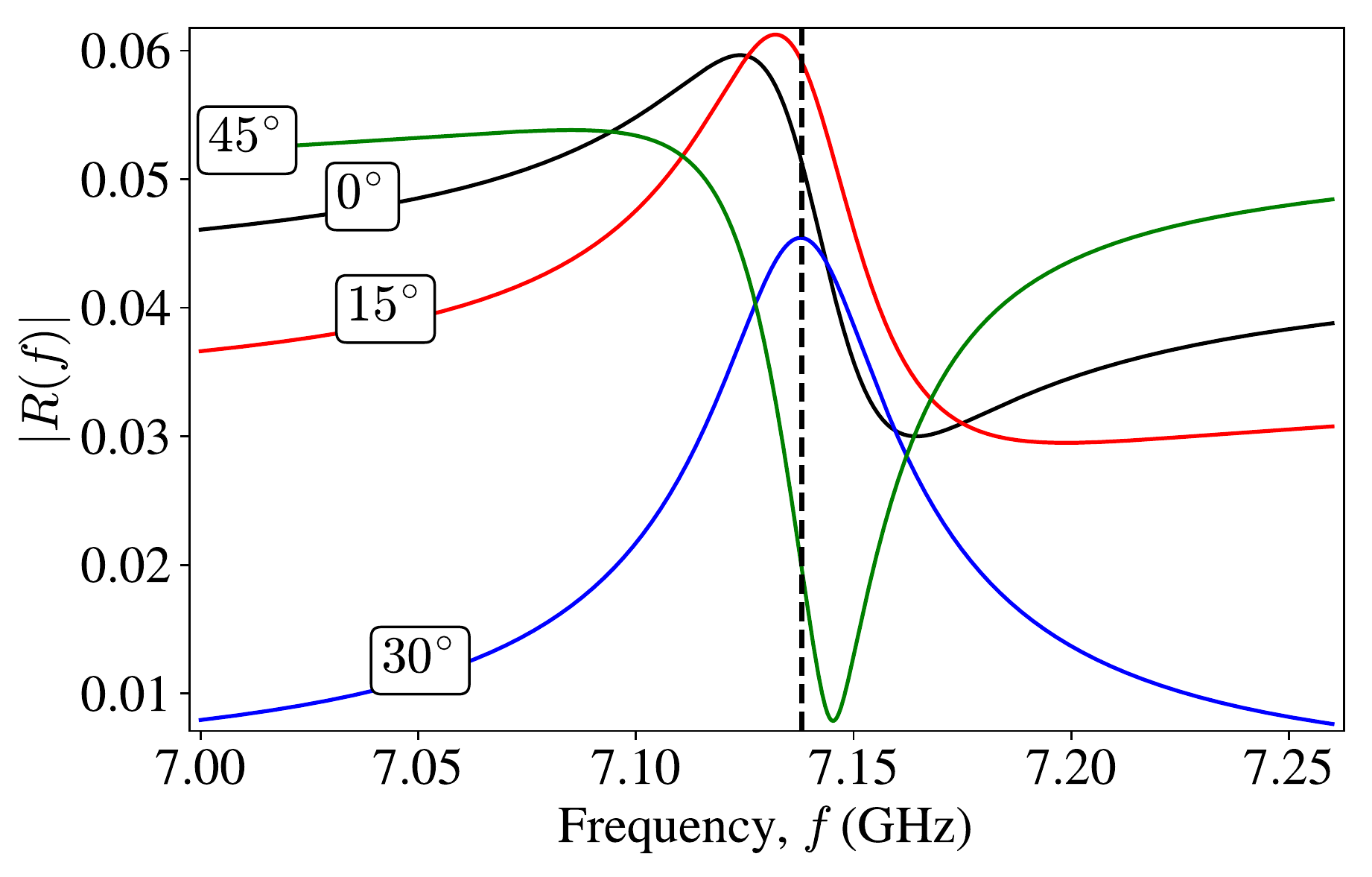}
    \caption{\label{fig:fig3}
    Peak $R(f)$ is enhanced and slightly shifted to the left in the oblique incidence geometry ($\theta > 0^{\circ}$). Coloured curves represent specific incidence angles sweeping from $0^{\circ}$ to $45^{\circ}$. Moderate Gilbert damping of $\alpha=10^{-3}$ is assumed. The dashed vertical line corresponds to the magnetoelastic resonance frequency.}
\end{figure}

\begin{align}
    R_{\omega} &= \frac{i\Gamma_{\mathrm{R}}/2}{(\omega-\omega_{\mathrm{ME}}) + i\Gamma_{\mathrm{R}}/2}e^{i\phi} + R_{0}, \\
    \phi&=-2\,\arctan\left[\frac{C}{C_{0}}\sqrt{\frac{\omega_{x}}{\omega_{y}}}\tan\theta\right], \nonumber
\end{align}\noindent 
where $R_{0}$ represents a smooth non-resonant contribution due to elastic mismatch at the interfaces, while $\phi$ represents a resonant phase, which is non-zero for finite $\theta$ and approaches~$\pi$ rapidly. In a system with no magnetic damping, the hybridization yields a resonance of  finite linewidth $\Gamma_{\mathrm{R}}$,
\begin{equation}
    \label{eqn:gammaR}
    \Gamma_{\mathrm{R}} = \frac{\gamma B^{2}}{2M_{\mathrm{s}}C^{2}\cos\theta}\sqrt{\rho_{0}C_{0}}\left(\omega_{y}\mathrm{cos}^{2}\theta+\frac{C^{2}}{C_{0}^{2}}\omega_{x}\mathrm{sin}^{2}\theta\right)\delta.
\end{equation}\noindent
The origin of this linewidth can be explained as follows. Due to the magneto-elastic coupling incident propagating acoustic modes can be converted into localised magnon modes. These modes in turn either decay due to the Gilbert damping or are re-emitted as phonons. The rates of these transitions are proportional to $\Gamma_{\mathrm{FMR}}$ and $\Gamma_{\mathrm{R}}$, respectively, and the total decay rate is $\Gamma = \Gamma_{\mathrm{R}}+\Gamma_{\mathrm{FMR}}$. This is similar to resonant scattering in quantum theory\cite{Landau_1965}, such that $\Gamma_{\mathrm{R}}$ and $\Gamma_{\mathrm{FMR}}$ are analogous to the the elastic $(\Gamma_{\mathrm{e}})$ and inelastic $(\Gamma_{\mathrm{i}})$ linewidths respectively. When $\alpha = 0$, $\Gamma_{\mathrm{FMR}}$ vanishes, and $\Gamma=\Gamma_{\mathrm{R}}$.

Acoustic waves in the geometry of Fig.~\ref{fig:fig1} can be scattered via several
channels. E.g. in a non-magnetic system ($B=0$), elastic mismatch can yield Fabry-P\'erot resonance due to the quarter wavelength matching of $\delta$ and the acoustic wavelength. However, this occurs at very high frequencies, which we do not consider here. To understand the resonant magneto-elastic response, it is instructive to consider first the case of normal incidence ($\theta=0$), when the demagnetising energy takes a simplified form due to the lack of immediate interfaces to form surface poles in $y$ the direction, so that $N_{x}=1$ and $N_{y}=0$. Including magneto-elastic coupling ($B\neq0$), we plot the frequency dependence of $R_{\omega}$ and $T_{\omega}$ using Eq.~(\ref{eqn:eqn11}) and ~(\ref{eqn:eqn12}) in Fig.\ref{fig:fig2}. To gain a quantitative insight, we analysed a magnetic inclusion made of cobalt ($\rho = 8900 \mathrm{kgm}^{-3}$, $B = 10 \mathrm{MPa}$, $C = 80 \mathrm{GPa}$, $\gamma = 176 \mathrm{GHz T^{-1}}$, $M = 1 \mathrm{MAm^{-1}}$), embedded into a non-magnetic matrix ($\rho_{0}  = 3192 \mathrm{kgm^{-3}}, C_0  = 298 \mathrm{GPa}$). To highlight the resonant behaviour, we first suppress $\alpha$ to $10^{-4}$. The reflection coefficient exhibits an asymmetric non-monotonic dependence, shown as a black curve in Fig.\ref{fig:fig2}(a), characteristic of Fano resonance.\cite{Graczyk_2017, Limonov_2017} This line shape can be attributed to coupling between the discrete FMR mode of the magnetic inclusion and the continuum of propagating acoustic modes in the surrounding non-magnetic material.\cite{Limonov_2017} If the two materials had matching elastic properties, $R_{\omega}$ would exhibit a symmetric Breit-Wigner lineshape.\cite{Landau_1965}
The transmission shown in Fig.\ref{fig:fig2}(b) exhibits an approximately symmetric dip near the resonance.\cite{Klaiman_2017} 
The absorbance $|A_{\omega}|^2=1-|R_{\omega}|^2-|T_{\omega}|^2$, 
shown in Fig.\ref{fig:fig2}(c) exhibits a symmetric peak, since the acoustic waves 
are damped in our model only due to the 
coupling with spin waves.

To consider how the magneto-elastic resonance is affected by the damping, we also plot the response for $\alpha$ of $10^{-3}$ and $10^{-2}$, red and blue curves in Fig.\ref{fig:fig2}, respectively. An increase of $\alpha$ from $10^{-4}$ to $10^{-3}$ significantly suppresses and broadens the resonant peak. For a more common, realistic value of $10^{-2}$ the resonance is quenched entirely.  A stronger magnetoelastic coupling (i.e. high values of $B$) could, in principle, countermand this suppression. This, however, is also likely to enhance the phonon contribution to the magnetic damping, leading to a correlation between $B$ and $\alpha$ observed in realistic magnetic materials.\cite{Emori_2017} 

To characterise the strength of the Fano resonance, we note that the fate of the magnon excited by the incident acoustic wave is decided by the relation between the emission rate $\Gamma_{\mathrm{R}}$, see Eq.~(\ref{eqn:gammaR}), and absorption rate $\Gamma_{\mathrm{FMR}}$.
Hence, we introduce the respective figure of merit as $\Upsilon = \Gamma_{\mathrm{R}}/\Gamma_{\mathrm{FMR}}$. 
This quantity depends upon the material parameters, device
geometry, and bias field. 
As seen from the first terms on the l.h.s. of Eqs.~(\ref{eqn:eqn6}) and (\ref{eqn:eqn7}), the relation between the dynamic magnetisation components~$m_{x, y}$ are determined by the quantities $\omega_{x}$ and $\omega_{y}$. 
Equating these terms, one finds $m_{x} \propto m_{y}\sqrt{\omega_{y}/\omega_{x}}$, i.e. the precession of $\bm{m}$ is highly elliptical,\cite{Kim_2012} 
due to the demagnetising field along~$x$.
This negatively affects the phonon-magnon 
coupling for normal incidence ($k_y = 0$): the acoustic  wave couples only to~$m_x$, as given by the second term in Eqs.~(\ref{eqn:eqn6}) and~(\ref{eqn:eqn7}). One way to mitigate this
is to increase $H_{\mathrm{B}}$,
moving the ratio $\omega_{y}/\omega_{x}$ closer to 1 and thus improving the figure of merit. To compare different magneto-elastic materials, the dependence on the layer thickness $\delta$ and elastic properties of the non-magnetic matrix (i.e. $\rho_0$ and $C_0$) can be eliminated by calculating a ratio of the figures of merit for the compared materials. The comparison can be performed either at the same value of the bias field, or at the same operating frequency.  The latter situation is more appropriate for a device application, but to avoid unphysical parameters, we present our results for the same $\mu_{0}H_{\mathrm{B}}$.  An example of such comparisons for yttrium iron garnet (YIG), cobalt (Co) and permalloy (Py) is offered in Table \ref{tab:table1}.  
\begin{table}[hb!]
\caption{\label{tab:table1}Comparison of the figure of merit $\Upsilon$ for different materials, assuming $\delta = 20$nm, $\mu_{0}H_{\mathrm{B}} = 50$mT and $C_{0} = 298$GPa.}
\begin{ruledtabular}
\begin{tabular}{cccc}
Parameters & YIG & Co & Py \\ 
\hline
$\Upsilon (\theta = 0^{\circ})$ & $4.3$x$10^{-2}$ & $1.7$x$10^{-3}$ & $2.7$x$10^{-4}$\\ 
$\Gamma_{\mathrm{R}}$  (ns$^{-1}$) &  $1.9$x$10^{-4}$ & $7.5$x$10^{-3}$ & $2.0$x$10^{-4}$ \\ 
$\Gamma_{\mathrm{FMR}}$  (ns$^{-1}$) & $4.4$x$10^{-3}$ & 4.3 & 0.74\\ 
\hline
$\Upsilon (\theta = 30^{\circ})$ & $4.1$x$10^{-2}$ & $2.5$x$10^{-3}$ & $2.8$x$10^{-4}$\\
$\Gamma_{\mathrm{R}}$  (ns$^{-1}$) &  $1.8$x$10^{-4}$ &  $1.1$x$10^{-2}$ &  $2.1$x$10^{-4}$\\ 
$\Gamma_{\mathrm{FMR}}$  (ns$^{-1}$) & $4.4$x$10^{-3}$ & 4.3 & 0.74\\
\hline
$f_{\mathrm{ME}} = \omega_{\mathrm{ME}}/2\pi$ (GHz) & 2.97 & 7.14 & 6.26\\ 
$B$ (MJm$^{-3}$) & 0.55 & 10 & -0.9\\ 
$C$ (GPa) & 74 & 80 & 50\\ 
$\rho$ (kgm$^{-3}$) & 5170 & 8900 & 8720\\ 
$\alpha$ & $0.9$x$10^{-4}$ & $1.8$x$10^{-2}$ & $4.0$x$10^{-3}$\\ 
$M_{\mathrm{s}}$ (kAm$^{-1}$) & 140 & 1000 & 760\\
\end{tabular}
\end{ruledtabular}
\end{table}

Another way to improve~$\Upsilon$ is to employ the oblique incidence ($\theta \neq 0$), in which the acoustic mode is also coupled to the magnetisation component~$m_y$. 
The latter is not suppressed by the demagnetisation effects if $N_y \ll 1$.
\begin{figure}[ht!]
    \centering
    \includegraphics[width=85mm]{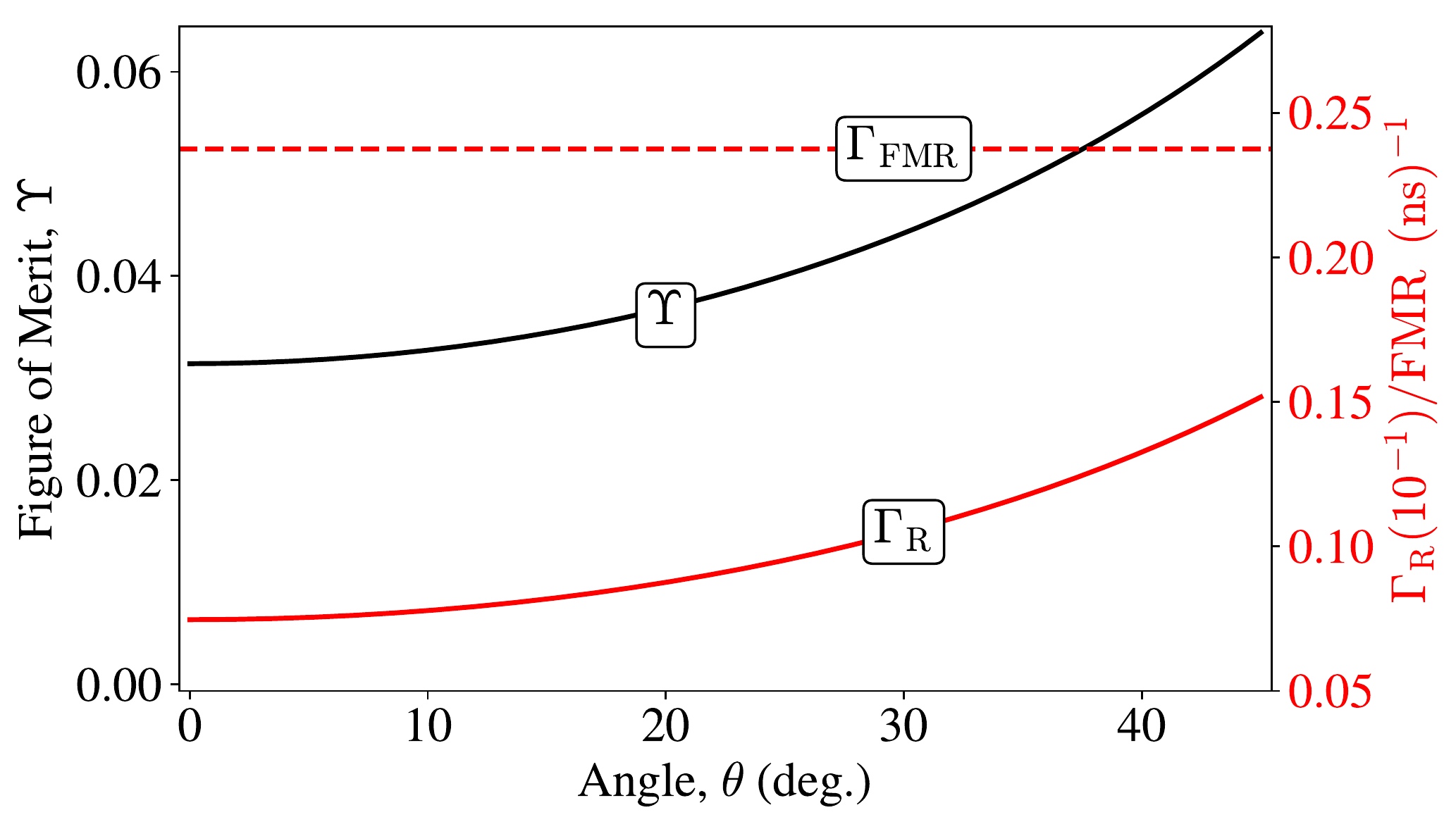}
    \caption{\label{fig:fig4}Figure of merit $\Upsilon$ and radiative linewidth $\Gamma_{\mathrm{R}}$ are both enhanced in the oblique incidence geometry ($\theta > 0^{\circ}$). Ferromagnetic linewidth $\Gamma_{\mathrm{FMR}}$ remains unchanged. Co is assumed with $\alpha = 10^{-3}.$}
\end{figure}
The resulting enhancement in~$\Upsilon$ is reflected in the full equation by the inclusion of $\omega_{x}$ and $\omega_{y}$ from $\Gamma_{\mathrm{R}}$,
\begin{equation}
    \Upsilon = \frac{\Gamma_{\mathrm{R}}}{\Gamma_{\mathrm{FMR}}} = \frac{\gamma \delta B^{2}}{2}\sqrt{\rho_{0}C_{0}}\frac{\left(H_{\mathrm{B}}\mathrm{cos}^2\theta+\frac{C^{2}}{C_{0}^{2}}M_{\mathrm{s}}\mathrm{sin}^2\theta\right)}{\alpha C^{2} M_{\mathrm{s}}^{2}\cos\theta},
\end{equation}
where $\omega_{x} \gg \omega_{y}$ and $H_{\mathrm{B}} \ll M_{\mathrm{s}}$ is assumed. For small $\theta$, the approximation $N_{x} \simeq 1$ and $N_{y} \simeq 0$ still holds. As a result, non-zero $\theta$ increases peak reflectivity, as seen in Fig.\ref{fig:fig3}. 
The evolution of the curves in Fig.3 with~$\theta$ is explained by the variation of the phase~$\phi$ of the resonant scattering relative to that of the non-resonant contribution~$R_0$. The latter changes its sign at incidence angle of about~$30^\circ$, which yields a nearly symmetric curve (blue), and an inverted Fano resonance at larger angles (green).  
Although larger incidence angles may be hard to implement in a practical device, the resonant scattering is still enhanced at smaller angles.

Above, we have focused on the simplest geometry that admits full analytic treatment. To implement our idea experimentally, particular care should be taken about the acoustic waves polarization and propagation direction relative to the direction of the magnetization.  Indeed, our choice maximises magnetoelastic response. If however, the polarization is orthogonal to the bias field $H_{\mathrm{B}}$,  i.e. $U_{z} = 0$, the coupling would be second-order in magnetization components $m_{x,y}$, and would not contribute to the linearized LLG equation. Furthermore, we have neglected the exchange and magneto-dipolar fields that could arise due to the non-uniformity of the magnetization. To assess the accuracy of this approximation, we note that the length scale of this non-uniformity is set by the acoustic wavelength $\lambda$, of about 420nm for our parameters rather than by the magnetic slab thickness $\delta$. The associated exchange field is $\mu_{0}M_{\mathrm{s}}(kl_{ex})^{2} \simeq 9$mT. The $k$-dependent contributions to the magneto-dipole field vanish at normal incidence but may become significant at oblique incidence, giving $\mu_{0}M_{\mathrm{s}}k_{y}\delta \simeq 98$mT at $\theta = 15^{\circ}$. In principle, these could increase the resonant frequency of the slab by a few GHz but would complicate the theory significantly. The detailed analysis of the associated effects is beyond the scope of this report.

In summary, we have demonstrated that the coupling between the magnetisation and strain fields can be used to control acoustic waves by magnetic inclusions. We show that the frequency dependence of the waves' reflection coefficient from the inclusions has a Fano-like lineshape, which is particularly sensitive to the magnetic damping.
Figure of merit is introduced to compare magnetoelastic materials and to
characterize device performance. In particular, the figure of merit
is significantly enhanced for oblique incidence of acoustic waves, which
enhances their coupling to the magnetic modes. 
We envision that further routes may be taken to transform our prototype designs into working devices, such as forming a magneto-acoustic metamaterial to take advantage of spatial resonance.

The research leading to these results has received funding from the Engineering and Physical Sciences Research Council of the United Kingdom (Grant No. EP/L015331/1) and from the European Union’s Horizon 2020 research and innovation program under Marie Skłodowska-Curie Grant Agreement No. 644348 (MagIC).  

\bibliography{MagnonPhononResonator}

\begin{thebibliography}{50}%
\makeatletter
\providecommand \@ifxundefined [1]{%
 \@ifx{#1\undefined}
}%
\providecommand \@ifnum [1]{%
 \ifnum #1\expandafter \@firstoftwo
 \else \expandafter \@secondoftwo
 \fi
}%
\providecommand \@ifx [1]{%
 \ifx #1\expandafter \@firstoftwo
 \else \expandafter \@secondoftwo
 \fi
}%
\providecommand \natexlab [1]{#1}%
\providecommand \enquote  [1]{``#1''}%
\providecommand \bibnamefont  [1]{#1}%
\providecommand \bibfnamefont [1]{#1}%
\providecommand \citenamefont [1]{#1}%
\providecommand \href@noop [0]{\@secondoftwo}%
\providecommand \href [0]{\begingroup \@sanitize@url \@href}%
\providecommand \@href[1]{\@@startlink{#1}\@@href}%
\providecommand \@@href[1]{\endgroup#1\@@endlink}%
\providecommand \@sanitize@url [0]{\catcode `\\12\catcode `\$12\catcode
  `\&12\catcode `\#12\catcode `\^12\catcode `\_12\catcode `\%12\relax}%
\providecommand \@@startlink[1]{}%
\providecommand \@@endlink[0]{}%
\providecommand \url  [0]{\begingroup\@sanitize@url \@url }%
\providecommand \@url [1]{\endgroup\@href {#1}{\urlprefix }}%
\providecommand \urlprefix  [0]{URL }%
\providecommand \Eprint [0]{\href }%
\providecommand \doibase [0]{http://dx.doi.org/}%
\providecommand \selectlanguage [0]{\@gobble}%
\providecommand \bibinfo  [0]{\@secondoftwo}%
\providecommand \bibfield  [0]{\@secondoftwo}%
\providecommand \translation [1]{[#1]}%
\providecommand \BibitemOpen [0]{}%
\providecommand \bibitemStop [0]{}%
\providecommand \bibitemNoStop [0]{.\EOS\space}%
\providecommand \EOS [0]{\spacefactor3000\relax}%
\providecommand \BibitemShut  [1]{\csname bibitem#1\endcsname}%
\let\auto@bib@innerbib\@empty
\bibitem [{\citenamefont {Feitelson}(1988)}]{Feitelson_1988}%
  \BibitemOpen
  \bibfield  {author} {\bibinfo {author} {\bibfnamefont {D.~G.}\ \bibnamefont
  {Feitelson}},\ }\href@noop {} {\emph {\bibinfo {title} {Optical computing: A
  survey for computer scientists}}}\ (\bibinfo  {publisher} {MIT Press},\
  \bibinfo {year} {1988})\BibitemShut {NoStop}%
\bibitem [{\citenamefont {Kruglyak}, \citenamefont {Demokritov},\ and\
  \citenamefont {Grundler}(2010)}]{Kruglyak_2010}%
  \BibitemOpen
  \bibfield  {author} {\bibinfo {author} {\bibfnamefont {V.~V.}\ \bibnamefont
  {Kruglyak}}, \bibinfo {author} {\bibfnamefont {S.~O.}\ \bibnamefont
  {Demokritov}}, \ and\ \bibinfo {author} {\bibfnamefont {D.}~\bibnamefont
  {Grundler}},\ }\bibfield  {title} {\enquote {\bibinfo {title} {Magnonics},}\
  }\href@noop {} {\bibfield  {journal} {\bibinfo  {journal} {J. Phys. D: Appl.
  Phys}\ }\textbf {\bibinfo {volume} {43}},\ \bibinfo {pages} {264001}
  (\bibinfo {year} {2010})}\BibitemShut {NoStop}%
\bibitem [{\citenamefont {Nikitov}\ \emph {et~al.}(2015)\citenamefont
  {Nikitov}, \citenamefont {Kalyabin}, \citenamefont {Lisenkov}, \citenamefont
  {Slavin}, \citenamefont {Barabanenkov}, \citenamefont {Osokin}, \citenamefont
  {Sadovnikov}, \citenamefont {Beginin}, \citenamefont {Morozova},
  \citenamefont {Sharaevsky}, \citenamefont {Filimonov}, \citenamefont
  {Khivintsev}, \citenamefont {Vysotsky}, \citenamefont {Sakharov},\ and\
  \citenamefont {Pavlov}}]{Nikitov_2015}%
  \BibitemOpen
  \bibfield  {author} {\bibinfo {author} {\bibfnamefont {S.}~\bibnamefont
  {Nikitov}}, \bibinfo {author} {\bibfnamefont {D.}~\bibnamefont {Kalyabin}},
  \bibinfo {author} {\bibfnamefont {I.}~\bibnamefont {Lisenkov}}, \bibinfo
  {author} {\bibfnamefont {A.}~\bibnamefont {Slavin}}, \bibinfo {author}
  {\bibfnamefont {Y.}~\bibnamefont {Barabanenkov}}, \bibinfo {author}
  {\bibfnamefont {S.}~\bibnamefont {Osokin}}, \bibinfo {author} {\bibfnamefont
  {A.}~\bibnamefont {Sadovnikov}}, \bibinfo {author} {\bibfnamefont
  {E.}~\bibnamefont {Beginin}}, \bibinfo {author} {\bibfnamefont
  {M.}~\bibnamefont {Morozova}}, \bibinfo {author} {\bibfnamefont
  {Y.}~\bibnamefont {Sharaevsky}}, \bibinfo {author} {\bibfnamefont
  {Y.}~\bibnamefont {Filimonov}}, \bibinfo {author} {\bibfnamefont
  {Y.}~\bibnamefont {Khivintsev}}, \bibinfo {author} {\bibfnamefont
  {S.}~\bibnamefont {Vysotsky}}, \bibinfo {author} {\bibfnamefont
  {V.}~\bibnamefont {Sakharov}}, \ and\ \bibinfo {author} {\bibfnamefont
  {E.}~\bibnamefont {Pavlov}},\ }\bibfield  {title} {\enquote {\bibinfo {title}
  {Magnonics: a new research area in spintronics and spin wave electronics},}\
  }\href@noop {} {\bibfield  {journal} {\bibinfo  {journal} {Phys. Uspekhi}\
  }\textbf {\bibinfo {volume} {58}},\ \bibinfo {pages} {1002} (\bibinfo {year}
  {2015})}\BibitemShut {NoStop}%
\bibitem [{\citenamefont {Akhiezer}, \citenamefont {yakhtar},\ and\
  \citenamefont {Peletminskii}(1968)}]{Akhiezer_1968}%
  \BibitemOpen
  \bibfield  {author} {\bibinfo {author} {\bibfnamefont {A.~I.}\ \bibnamefont
  {Akhiezer}}, \bibinfo {author} {\bibfnamefont {V.~G.~B.}\ \bibnamefont
  {yakhtar}}, \ and\ \bibinfo {author} {\bibfnamefont {S.~V.}\ \bibnamefont
  {Peletminskii}},\ }\href@noop {} {\emph {\bibinfo {title} {Spin waves}}}\
  (\bibinfo  {publisher} {North-Holland, Amsterdam},\ \bibinfo {year}
  {1968})\BibitemShut {NoStop}%
\bibitem [{\citenamefont {Krivoruchko}(2015)}]{Krivoruchko_2015}%
  \BibitemOpen
  \bibfield  {author} {\bibinfo {author} {\bibfnamefont {V.~N.}\ \bibnamefont
  {Krivoruchko}},\ }\bibfield  {title} {\enquote {\bibinfo {title} {Spin waves
  damping in nanometre-scale magnetic materials (review article)},}\
  }\href@noop {} {\bibfield  {journal} {\bibinfo  {journal} {Low Temp. Phys.}\
  }\textbf {\bibinfo {volume} {41}},\ \bibinfo {pages} {670} (\bibinfo {year}
  {2015})}\BibitemShut {NoStop}%
\bibitem [{\citenamefont {Azzawi}, \citenamefont {Hindmarch},\ and\
  \citenamefont {Atkinson}(2017)}]{Azzawi_2017}%
  \BibitemOpen
  \bibfield  {author} {\bibinfo {author} {\bibfnamefont {S.}~\bibnamefont
  {Azzawi}}, \bibinfo {author} {\bibfnamefont {A.~T.}\ \bibnamefont
  {Hindmarch}}, \ and\ \bibinfo {author} {\bibfnamefont {D.}~\bibnamefont
  {Atkinson}},\ }\bibfield  {title} {\enquote {\bibinfo {title} {Magnetic
  damping phenomena in ferromagnetic thin-films and multilayers},}\ }\href@noop
  {} {\bibfield  {journal} {\bibinfo  {journal} {J. Phys. D: Appl. Phys}\
  }\textbf {\bibinfo {volume} {50}},\ \bibinfo {pages} {471001} (\bibinfo
  {year} {2017})}\BibitemShut {NoStop}%
\bibitem [{\citenamefont {Collins}(1984)}]{Collins_1984}%
  \BibitemOpen
  \bibfield  {author} {\bibinfo {author} {\bibfnamefont {J.~H.}\ \bibnamefont
  {Collins}},\ }\bibfield  {title} {\enquote {\bibinfo {title} {Short history
  of microwave acoustics},}\ }\href@noop {} {\bibfield  {journal} {\bibinfo
  {journal} {IEEE Trans. Microwave Theory Tech.}\ }\textbf {\bibinfo {volume}
  {MTT-32}},\ \bibinfo {pages} {1127} (\bibinfo {year} {1984})}\BibitemShut
  {NoStop}%
\bibitem [{\citenamefont {Kittel}(1958)}]{Kittel_1958}%
  \BibitemOpen
  \bibfield  {author} {\bibinfo {author} {\bibfnamefont {C.}~\bibnamefont
  {Kittel}},\ }\bibfield  {title} {\enquote {\bibinfo {title} {Interaction of
  spin waves and ultrasonic waves in ferromagnetic crystals},}\ }\href@noop {}
  {\bibfield  {journal} {\bibinfo  {journal} {Phys. Rev.}\ }\textbf {\bibinfo
  {volume} {110}},\ \bibinfo {pages} {836} (\bibinfo {year}
  {1958})}\BibitemShut {NoStop}%
\bibitem [{\citenamefont {B\"ommel}\ and\ \citenamefont
  {Dransfeld}(1959)}]{Bommel_1959}%
  \BibitemOpen
  \bibfield  {author} {\bibinfo {author} {\bibfnamefont {H.}~\bibnamefont
  {B\"ommel}}\ and\ \bibinfo {author} {\bibfnamefont {K.}~\bibnamefont
  {Dransfeld}},\ }\bibfield  {title} {\enquote {\bibinfo {title} {Excitation of
  hypersonic waves by ferromagnetic resonance},}\ }\href@noop {} {\bibfield
  {journal} {\bibinfo  {journal} {Phys. Rev. Lett}\ }\textbf {\bibinfo {volume}
  {3}},\ \bibinfo {pages} {83} (\bibinfo {year} {1959})}\BibitemShut {NoStop}%
\bibitem [{\citenamefont {Dreher}\ \emph {et~al.}(2012)\citenamefont {Dreher},
  \citenamefont {Weiler}, \citenamefont {Pernpeintner}, \citenamefont {Huebl},
  \citenamefont {Gross}, \citenamefont {Brandt},\ and\ \citenamefont
  {Goennenwein}}]{Dreher_2012}%
  \BibitemOpen
  \bibfield  {author} {\bibinfo {author} {\bibfnamefont {L.}~\bibnamefont
  {Dreher}}, \bibinfo {author} {\bibfnamefont {M.}~\bibnamefont {Weiler}},
  \bibinfo {author} {\bibfnamefont {M.}~\bibnamefont {Pernpeintner}}, \bibinfo
  {author} {\bibfnamefont {H.}~\bibnamefont {Huebl}}, \bibinfo {author}
  {\bibfnamefont {R.}~\bibnamefont {Gross}}, \bibinfo {author} {\bibfnamefont
  {M.}~\bibnamefont {Brandt}}, \ and\ \bibinfo {author} {\bibfnamefont
  {S.}~\bibnamefont {Goennenwein}},\ }\bibfield  {title} {\enquote {\bibinfo
  {title} {Surface acoustic wave driven ferromagnetic resonance in nickel thin
  films: Theory and experiment},}\ }\href@noop {} {\bibfield  {journal}
  {\bibinfo  {journal} {Phys. Rev. B}\ }\textbf {\bibinfo {volume} {86}},\
  \bibinfo {pages} {134415} (\bibinfo {year} {2012})}\BibitemShut {NoStop}%
\bibitem [{\citenamefont {Li}\ \emph {et~al.}(2012)\citenamefont {Li},
  \citenamefont {Ren}, \citenamefont {Wang}, \citenamefont {Zhang},
  \citenamefont {Hanggi},\ and\ \citenamefont {Li}}]{Li_2012}%
  \BibitemOpen
  \bibfield  {author} {\bibinfo {author} {\bibfnamefont {N.}~\bibnamefont
  {Li}}, \bibinfo {author} {\bibfnamefont {J.}~\bibnamefont {Ren}}, \bibinfo
  {author} {\bibfnamefont {L.}~\bibnamefont {Wang}}, \bibinfo {author}
  {\bibfnamefont {G.}~\bibnamefont {Zhang}}, \bibinfo {author} {\bibfnamefont
  {P.}~\bibnamefont {Hanggi}}, \ and\ \bibinfo {author} {\bibfnamefont {B.~W.}\
  \bibnamefont {Li}},\ }\bibfield  {title} {\enquote {\bibinfo {title}
  {Colloquium: Phononics: Manipulating heat flow with electronic analogs and
  beyond},}\ }\href@noop {} {\bibfield  {journal} {\bibinfo  {journal} {Rev.
  Mod. Phys.}\ }\textbf {\bibinfo {volume} {84}},\ \bibinfo {pages} {1045}
  (\bibinfo {year} {2012})}\BibitemShut {NoStop}%
\bibitem [{\citenamefont {Maldovan}(2013)}]{Maldovan_2013}%
  \BibitemOpen
  \bibfield  {author} {\bibinfo {author} {\bibfnamefont {M.}~\bibnamefont
  {Maldovan}},\ }\bibfield  {title} {\enquote {\bibinfo {title} {Sound and heat
  revolutions in phononics},}\ }\href@noop {} {\bibfield  {journal} {\bibinfo
  {journal} {Nature}\ }\textbf {\bibinfo {volume} {503}},\ \bibinfo {pages}
  {209} (\bibinfo {year} {2013})}\BibitemShut {NoStop}%
\bibitem [{\citenamefont {H\"ollander}\ \emph {et~al.}(2018)\citenamefont
  {H\"ollander}, \citenamefont {M\"uller}, \citenamefont {Schmalz},
  \citenamefont {Gerken},\ and\ \citenamefont {McCord}}]{Hollander_2018}%
  \BibitemOpen
  \bibfield  {author} {\bibinfo {author} {\bibfnamefont {R.~B.}\ \bibnamefont
  {H\"ollander}}, \bibinfo {author} {\bibfnamefont {C.}~\bibnamefont
  {M\"uller}}, \bibinfo {author} {\bibfnamefont {J.}~\bibnamefont {Schmalz}},
  \bibinfo {author} {\bibfnamefont {M.}~\bibnamefont {Gerken}}, \ and\ \bibinfo
  {author} {\bibfnamefont {J.}~\bibnamefont {McCord}},\ }\bibfield  {title}
  {\enquote {\bibinfo {title} {Magnetic domain walls as broadband spin wave and
  elastic magnetisation wave emitters},}\ }\href@noop {} {\bibfield  {journal}
  {\bibinfo  {journal} {Sci. Rep}\ }\textbf {\bibinfo {volume} {8}},\ \bibinfo
  {pages} {13871} (\bibinfo {year} {2018})}\BibitemShut {NoStop}%
\bibitem [{\citenamefont {Streib}, \citenamefont {Keshtgar},\ and\
  \citenamefont {Bauer}(2018)}]{Bauer_2018}%
  \BibitemOpen
  \bibfield  {author} {\bibinfo {author} {\bibfnamefont {S.}~\bibnamefont
  {Streib}}, \bibinfo {author} {\bibfnamefont {H.}~\bibnamefont {Keshtgar}}, \
  and\ \bibinfo {author} {\bibfnamefont {G.~E.~W.}\ \bibnamefont {Bauer}},\
  }\bibfield  {title} {\enquote {\bibinfo {title} {Damping of magnetization
  dynamics by phonon pumping},}\ }\href@noop {} {\bibfield  {journal} {\bibinfo
   {journal} {Phys. Rev. Lett}\ }\textbf {\bibinfo {volume} {121}},\ \bibinfo
  {pages} {027202} (\bibinfo {year} {2018})}\BibitemShut {NoStop}%
\bibitem [{\citenamefont {Thingstad}\ \emph {et~al.}(2019)\citenamefont
  {Thingstad}, \citenamefont {Kamra}, \citenamefont {Brataas},\ and\
  \citenamefont {Sudb\o}}]{Thingstad_2019}%
  \BibitemOpen
  \bibfield  {author} {\bibinfo {author} {\bibfnamefont {E.}~\bibnamefont
  {Thingstad}}, \bibinfo {author} {\bibfnamefont {A.}~\bibnamefont {Kamra}},
  \bibinfo {author} {\bibfnamefont {A.}~\bibnamefont {Brataas}}, \ and\
  \bibinfo {author} {\bibfnamefont {A.}~\bibnamefont {Sudb\o}},\ }\bibfield
  {title} {\enquote {\bibinfo {title} {Chiral phonon transport induced by
  topological magnons},}\ }\href@noop {} {\bibfield  {journal} {\bibinfo
  {journal} {Phys. Rev. Lett}\ }\textbf {\bibinfo {volume} {122}},\ \bibinfo
  {pages} {107201} (\bibinfo {year} {2019})}\BibitemShut {NoStop}%
\bibitem [{\citenamefont {Li}\ \emph {et~al.}(2017)\citenamefont {Li},
  \citenamefont {Labanowski}, \citenamefont {Salahuddin},\ and\ \citenamefont
  {Lynch}}]{Li_2017}%
  \BibitemOpen
  \bibfield  {author} {\bibinfo {author} {\bibfnamefont {X.}~\bibnamefont
  {Li}}, \bibinfo {author} {\bibfnamefont {D.}~\bibnamefont {Labanowski}},
  \bibinfo {author} {\bibfnamefont {S.}~\bibnamefont {Salahuddin}}, \ and\
  \bibinfo {author} {\bibfnamefont {C.~S.}\ \bibnamefont {Lynch}},\ }\bibfield
  {title} {\enquote {\bibinfo {title} {Spin wave generation by surface acoustic
  waves},}\ }\href@noop {} {\bibfield  {journal} {\bibinfo  {journal} {J. Appl.
  Phys.}\ }\textbf {\bibinfo {volume} {122}},\ \bibinfo {pages} {043904}
  (\bibinfo {year} {2017})}\BibitemShut {NoStop}%
\bibitem [{\citenamefont {Gowtham}\ \emph {et~al.}(2015)\citenamefont
  {Gowtham}, \citenamefont {Moriyama}, \citenamefont {Ralph},\ and\
  \citenamefont {Buhrman}}]{Gowtham_2015}%
  \BibitemOpen
  \bibfield  {author} {\bibinfo {author} {\bibfnamefont {P.~G.}\ \bibnamefont
  {Gowtham}}, \bibinfo {author} {\bibfnamefont {T.}~\bibnamefont {Moriyama}},
  \bibinfo {author} {\bibfnamefont {D.~C.}\ \bibnamefont {Ralph}}, \ and\
  \bibinfo {author} {\bibfnamefont {R.~A.}\ \bibnamefont {Buhrman}},\
  }\bibfield  {title} {\enquote {\bibinfo {title} {Traveling surface spin-wave
  resonance spectroscopy using surface acoustic waves},}\ }\href@noop {}
  {\bibfield  {journal} {\bibinfo  {journal} {J. Appl. Phys.}\ }\textbf
  {\bibinfo {volume} {118}},\ \bibinfo {pages} {233910} (\bibinfo {year}
  {2015})}\BibitemShut {NoStop}%
\bibitem [{\citenamefont {Ulrichs}\ \emph {et~al.}(2017)\citenamefont
  {Ulrichs}, \citenamefont {Meyer}, \citenamefont {D\"oring}, \citenamefont
  {Eberl},\ and\ \citenamefont {Krebs}}]{Ulrichs_2017}%
  \BibitemOpen
  \bibfield  {author} {\bibinfo {author} {\bibfnamefont {H.}~\bibnamefont
  {Ulrichs}}, \bibinfo {author} {\bibfnamefont {D.}~\bibnamefont {Meyer}},
  \bibinfo {author} {\bibfnamefont {F.}~\bibnamefont {D\"oring}}, \bibinfo
  {author} {\bibfnamefont {C.}~\bibnamefont {Eberl}}, \ and\ \bibinfo {author}
  {\bibfnamefont {H.~U.}\ \bibnamefont {Krebs}},\ }\bibfield  {title} {\enquote
  {\bibinfo {title} {Spectral control of elastic dynamics in metallic
  nano-cavities},}\ }\href@noop {} {\bibfield  {journal} {\bibinfo  {journal}
  {Sci. Rep.}\ }\textbf {\bibinfo {volume} {7}},\ \bibinfo {pages} {10600}
  (\bibinfo {year} {2017})}\BibitemShut {NoStop}%
\bibitem [{\citenamefont {Gurevich}(1965)}]{Gurevich_1965}%
  \BibitemOpen
  \bibfield  {author} {\bibinfo {author} {\bibfnamefont {A.~G.}\ \bibnamefont
  {Gurevich}},\ }\bibfield  {title} {\enquote {\bibinfo {title} {Parametric
  amplification of magnetic waves in ferrites by an elastic wave},}\
  }\href@noop {} {\bibfield  {journal} {\bibinfo  {journal} {Phys. Sol. State}\
  }\textbf {\bibinfo {volume} {6}},\ \bibinfo {pages} {1885} (\bibinfo {year}
  {1965})}\BibitemShut {NoStop}%
\bibitem [{\citenamefont {Keshtgar}, \citenamefont {Zareyan},\ and\
  \citenamefont {Bauer}(2014)}]{Keshtgar_2014}%
  \BibitemOpen
  \bibfield  {author} {\bibinfo {author} {\bibfnamefont {H.}~\bibnamefont
  {Keshtgar}}, \bibinfo {author} {\bibfnamefont {M.}~\bibnamefont {Zareyan}}, \
  and\ \bibinfo {author} {\bibfnamefont {G.~E.~W.}\ \bibnamefont {Bauer}},\
  }\bibfield  {title} {\enquote {\bibinfo {title} {Acoustic parametric pumping
  of spin wave},}\ }\href@noop {} {\bibfield  {journal} {\bibinfo  {journal}
  {Sol. State. Commun.}\ }\textbf {\bibinfo {volume} {198}},\ \bibinfo {pages}
  {30} (\bibinfo {year} {2014})}\BibitemShut {NoStop}%
\bibitem [{\citenamefont {Chowdhury}, \citenamefont {Dhagat},\ and\
  \citenamefont {Jander}(2015)}]{Chowdhury_2015}%
  \BibitemOpen
  \bibfield  {author} {\bibinfo {author} {\bibfnamefont {P.}~\bibnamefont
  {Chowdhury}}, \bibinfo {author} {\bibfnamefont {P.}~\bibnamefont {Dhagat}}, \
  and\ \bibinfo {author} {\bibfnamefont {A.}~\bibnamefont {Jander}},\
  }\bibfield  {title} {\enquote {\bibinfo {title} {Parametric amplification of
  spin waves using bulk acoustic waves},}\ }\href@noop {} {\bibfield  {journal}
  {\bibinfo  {journal} {IEEE Trans. Magn.}\ }\textbf {\bibinfo {volume} {51}},\
  \bibinfo {pages} {1300904} (\bibinfo {year} {2015})}\BibitemShut {NoStop}%
\bibitem [{\citenamefont {Litvinenko}\ \emph {et~al.}(2015)\citenamefont
  {Litvinenko}, \citenamefont {Sadovnikov}, \citenamefont {Tikhonov},\ and\
  \citenamefont {Nikitov}}]{Litvinenko_2015}%
  \BibitemOpen
  \bibfield  {author} {\bibinfo {author} {\bibfnamefont {A.~N.}\ \bibnamefont
  {Litvinenko}}, \bibinfo {author} {\bibfnamefont {A.~V.}\ \bibnamefont
  {Sadovnikov}}, \bibinfo {author} {\bibfnamefont {V.~V.}\ \bibnamefont
  {Tikhonov}}, \ and\ \bibinfo {author} {\bibfnamefont {S.~A.}\ \bibnamefont
  {Nikitov}},\ }\bibfield  {title} {\enquote {\bibinfo {title} {Brillouin light
  scattering spectroscopy of magneto-acoustic resonances in a thin-film garnet
  resonator},}\ }\href@noop {} {\bibfield  {journal} {\bibinfo  {journal} {IEEE
  Magn. Lett.}\ }\textbf {\bibinfo {volume} {6}},\ \bibinfo {pages} {3200204}
  (\bibinfo {year} {2015})}\BibitemShut {NoStop}%
\bibitem [{\citenamefont {Zhang}\ \emph {et~al.}(2016)\citenamefont {Zhang},
  \citenamefont {Zou}, \citenamefont {Jiang},\ and\ \citenamefont
  {Tang}}]{Zhango_2016}%
  \BibitemOpen
  \bibfield  {author} {\bibinfo {author} {\bibfnamefont {X.~F.}\ \bibnamefont
  {Zhang}}, \bibinfo {author} {\bibfnamefont {C.~L.}\ \bibnamefont {Zou}},
  \bibinfo {author} {\bibfnamefont {L.}~\bibnamefont {Jiang}}, \ and\ \bibinfo
  {author} {\bibfnamefont {H.~X.}\ \bibnamefont {Tang}},\ }\bibfield  {title}
  {\enquote {\bibinfo {title} {Cavity magnomechanics},}\ }\href@noop {}
  {\bibfield  {journal} {\bibinfo  {journal} {Sci. Adv.}\ }\textbf {\bibinfo
  {volume} {2}},\ \bibinfo {pages} {e1501286} (\bibinfo {year}
  {2016})}\BibitemShut {NoStop}%
\bibitem [{\citenamefont {Kong}\ \emph {et~al.}(2019)\citenamefont {Kong},
  \citenamefont {Wang}, \citenamefont {Liu}, \citenamefont {Xiong},\ and\
  \citenamefont {Wu}}]{Kong_2019}%
  \BibitemOpen
  \bibfield  {author} {\bibinfo {author} {\bibfnamefont {C.}~\bibnamefont
  {Kong}}, \bibinfo {author} {\bibfnamefont {B.}~\bibnamefont {Wang}}, \bibinfo
  {author} {\bibfnamefont {Z.~X.}\ \bibnamefont {Liu}}, \bibinfo {author}
  {\bibfnamefont {H.}~\bibnamefont {Xiong}}, \ and\ \bibinfo {author}
  {\bibfnamefont {Y.}~\bibnamefont {Wu}},\ }\bibfield  {title} {\enquote
  {\bibinfo {title} {Magnetically controllable slow light based on
  magnetostrictive forces},}\ }\href@noop {} {\bibfield  {journal} {\bibinfo
  {journal} {Opt. Express}\ }\textbf {\bibinfo {volume} {27}},\ \bibinfo
  {pages} {5544} (\bibinfo {year} {2019})}\BibitemShut {NoStop}%
\bibitem [{\citenamefont {Wang}\ and\ \citenamefont {lin
  Hsu}(1970)}]{Wang_1970}%
  \BibitemOpen
  \bibfield  {author} {\bibinfo {author} {\bibfnamefont {S.}~\bibnamefont
  {Wang}}\ and\ \bibinfo {author} {\bibfnamefont {T.}~\bibnamefont {lin Hsu}},\
  }\bibfield  {title} {\enquote {\bibinfo {title} {Observation of spin-wave
  spectrum in an instability and mode-locking experiment},}\ }\href@noop {}
  {\bibfield  {journal} {\bibinfo  {journal} {Appl. Phys. Lett}\ }\textbf
  {\bibinfo {volume} {16}},\ \bibinfo {pages} {534} (\bibinfo {year}
  {1970})}\BibitemShut {NoStop}%
\bibitem [{\citenamefont {Nikitov}\ \emph {et~al.}(2012)\citenamefont
  {Nikitov}, \citenamefont {Filimonov}, \citenamefont {Vysotsky}, \citenamefont
  {Khivintsev},\ and\ \citenamefont {Pavlov}}]{Nikitov_2012}%
  \BibitemOpen
  \bibfield  {author} {\bibinfo {author} {\bibfnamefont {S.}~\bibnamefont
  {Nikitov}}, \bibinfo {author} {\bibfnamefont {Y.}~\bibnamefont {Filimonov}},
  \bibinfo {author} {\bibfnamefont {S.}~\bibnamefont {Vysotsky}}, \bibinfo
  {author} {\bibfnamefont {Y.}~\bibnamefont {Khivintsev}}, \ and\ \bibinfo
  {author} {\bibfnamefont {E.}~\bibnamefont {Pavlov}},\ }\bibfield  {title}
  {\enquote {\bibinfo {title} {Yttrium iron garnet based phononic-magnonic
  crystal},}\ }\href@noop {} {\bibfield  {journal} {\bibinfo  {journal} {Proc.
  of 2012 IEEE International Ultrasonics Symposium}\ ,\ \bibinfo {pages}
  {1240--1243}} (\bibinfo {year} {2012})}\BibitemShut {NoStop}%
\bibitem [{\citenamefont {Graczyk}, \citenamefont {K\l{}os},\ and\
  \citenamefont {Krawczyk}(2017)}]{Graczyk_2017}%
  \BibitemOpen
  \bibfield  {author} {\bibinfo {author} {\bibfnamefont {P.}~\bibnamefont
  {Graczyk}}, \bibinfo {author} {\bibfnamefont {J.}~\bibnamefont {K\l{}os}}, \
  and\ \bibinfo {author} {\bibfnamefont {M.}~\bibnamefont {Krawczyk}},\
  }\bibfield  {title} {\enquote {\bibinfo {title} {Broadband magnetoelastic
  coupling in magnonic-phononic crystals for high-frequency nanoscale spin-wave
  generation},}\ }\href@noop {} {\bibfield  {journal} {\bibinfo  {journal}
  {Phys. Rev. B}\ }\textbf {\bibinfo {volume} {95}},\ \bibinfo {pages} {104425}
  (\bibinfo {year} {2017})}\BibitemShut {NoStop}%
\bibitem [{\citenamefont {Chumak}\ \emph {et~al.}(2010)\citenamefont {Chumak},
  \citenamefont {Dhagat}, \citenamefont {Jander}, \citenamefont {Serga},\ and\
  \citenamefont {Hillebrands}}]{Chumak_2010}%
  \BibitemOpen
  \bibfield  {author} {\bibinfo {author} {\bibfnamefont {A.~V.}\ \bibnamefont
  {Chumak}}, \bibinfo {author} {\bibfnamefont {P.}~\bibnamefont {Dhagat}},
  \bibinfo {author} {\bibfnamefont {A.}~\bibnamefont {Jander}}, \bibinfo
  {author} {\bibfnamefont {A.~A.}\ \bibnamefont {Serga}}, \ and\ \bibinfo
  {author} {\bibfnamefont {B.}~\bibnamefont {Hillebrands}},\ }\bibfield
  {title} {\enquote {\bibinfo {title} {Reverse doppler effect of magnons with
  negative group velocity scattered from a moving bragg grating},}\ }\href@noop
  {} {\bibfield  {journal} {\bibinfo  {journal} {Phys. Rev. B}\ }\textbf
  {\bibinfo {volume} {81}},\ \bibinfo {pages} {140404} (\bibinfo {year}
  {2010})}\BibitemShut {NoStop}%
\bibitem [{\citenamefont {Kryshtal}\ and\ \citenamefont
  {Medved}(2017{\natexlab{a}})}]{Kryshtal_2017_1}%
  \BibitemOpen
  \bibfield  {author} {\bibinfo {author} {\bibfnamefont {R.~G.}\ \bibnamefont
  {Kryshtal}}\ and\ \bibinfo {author} {\bibfnamefont {A.~V.}\ \bibnamefont
  {Medved}},\ }\bibfield  {title} {\enquote {\bibinfo {title} {Influence of
  magnetic anisotropy on dynamic magnonic crystals created by surface acoustic
  waves in yttrium iron garnet films},}\ }\href@noop {} {\bibfield  {journal}
  {\bibinfo  {journal} {J. Magn. Magn. Mater.}\ }\textbf {\bibinfo {volume}
  {426}},\ \bibinfo {pages} {666} (\bibinfo {year}
  {2017}{\natexlab{a}})}\BibitemShut {NoStop}%
\bibitem [{\citenamefont {Kryshtal}\ and\ \citenamefont
  {Medved}(2017{\natexlab{b}})}]{Kryshtal_2017_2}%
  \BibitemOpen
  \bibfield  {author} {\bibinfo {author} {\bibfnamefont {R.~G.}\ \bibnamefont
  {Kryshtal}}\ and\ \bibinfo {author} {\bibfnamefont {A.~V.}\ \bibnamefont
  {Medved}},\ }\bibfield  {title} {\enquote {\bibinfo {title} {Nonlinear spin
  waves in dynamic magnonic crystals created by surface acoustic waves in
  yttrium iron garnet films},}\ }\href@noop {} {\bibfield  {journal} {\bibinfo
  {journal} {J. Phys. D: Appl. Phys.}\ }\textbf {\bibinfo {volume} {50}},\
  \bibinfo {pages} {495004} (\bibinfo {year} {2017}{\natexlab{b}})}\BibitemShut
  {NoStop}%
\bibitem [{\citenamefont {Takahashi}\ and\ \citenamefont
  {Nagaosa}(2016)}]{Takahashi_2016}%
  \BibitemOpen
  \bibfield  {author} {\bibinfo {author} {\bibfnamefont {R.}~\bibnamefont
  {Takahashi}}\ and\ \bibinfo {author} {\bibfnamefont {N.}~\bibnamefont
  {Nagaosa}},\ }\bibfield  {title} {\enquote {\bibinfo {title} {Berry curvature
  in magnon-phonon hybrid systems},}\ }\href@noop {} {\bibfield  {journal}
  {\bibinfo  {journal} {Sci. Rep.}\ }\textbf {\bibinfo {volume} {117}},\
  \bibinfo {pages} {217205} (\bibinfo {year} {2016})}\BibitemShut {NoStop}%
\bibitem [{\citenamefont {Uchida}\ \emph {et~al.}(2011)\citenamefont {Uchida},
  \citenamefont {Adachi}, \citenamefont {An}, \citenamefont {Ota},
  \citenamefont {Toda}, \citenamefont {Hillebrands}, \citenamefont {Maekawa},\
  and\ \citenamefont {Saitoh}}]{Uchida_2011}%
  \BibitemOpen
  \bibfield  {author} {\bibinfo {author} {\bibfnamefont {K.}~\bibnamefont
  {Uchida}}, \bibinfo {author} {\bibfnamefont {H.}~\bibnamefont {Adachi}},
  \bibinfo {author} {\bibfnamefont {T.}~\bibnamefont {An}}, \bibinfo {author}
  {\bibfnamefont {T.}~\bibnamefont {Ota}}, \bibinfo {author} {\bibfnamefont
  {M.}~\bibnamefont {Toda}}, \bibinfo {author} {\bibfnamefont {B.}~\bibnamefont
  {Hillebrands}}, \bibinfo {author} {\bibfnamefont {S.}~\bibnamefont
  {Maekawa}}, \ and\ \bibinfo {author} {\bibfnamefont {E.}~\bibnamefont
  {Saitoh}},\ }\bibfield  {title} {\enquote {\bibinfo {title} {Long-range spin
  seebeck effect and acoustic spin pumping},}\ }\href@noop {} {\bibfield
  {journal} {\bibinfo  {journal} {Nature Mater.}\ }\textbf {\bibinfo {volume}
  {10}},\ \bibinfo {pages} {737} (\bibinfo {year} {2011})}\BibitemShut
  {NoStop}%
\bibitem [{\citenamefont {Polzikova}\ \emph {et~al.}(2018)\citenamefont
  {Polzikova}, \citenamefont {Alekseev}, \citenamefont {Luzanov},\ and\
  \citenamefont {Raevskiy}}]{Polzikova_2018}%
  \BibitemOpen
  \bibfield  {author} {\bibinfo {author} {\bibfnamefont {N.~I.}\ \bibnamefont
  {Polzikova}}, \bibinfo {author} {\bibfnamefont {S.~G.}\ \bibnamefont
  {Alekseev}}, \bibinfo {author} {\bibfnamefont {V.~A.}\ \bibnamefont
  {Luzanov}}, \ and\ \bibinfo {author} {\bibfnamefont {A.~O.}\ \bibnamefont
  {Raevskiy}},\ }\bibfield  {title} {\enquote {\bibinfo {title}
  {Electroacoustic excitation of spin waves and their detection due to the
  inverse spin hall effect},}\ }\href@noop {} {\bibfield  {journal} {\bibinfo
  {journal} {Phys. Solid State}\ }\textbf {\bibinfo {volume} {60}},\ \bibinfo
  {pages} {2211} (\bibinfo {year} {2018})}\BibitemShut {NoStop}%
\bibitem [{\citenamefont {Yahagi}\ \emph {et~al.}(2014)\citenamefont {Yahagi},
  \citenamefont {Harteneck}, \citenamefont {Cabrini},\ and\ \citenamefont
  {Schmidt}}]{Yahagi_2014}%
  \BibitemOpen
  \bibfield  {author} {\bibinfo {author} {\bibfnamefont {Y.}~\bibnamefont
  {Yahagi}}, \bibinfo {author} {\bibfnamefont {B.}~\bibnamefont {Harteneck}},
  \bibinfo {author} {\bibfnamefont {S.}~\bibnamefont {Cabrini}}, \ and\
  \bibinfo {author} {\bibfnamefont {H.}~\bibnamefont {Schmidt}},\ }\bibfield
  {title} {\enquote {\bibinfo {title} {Controlling nanomagnet magnetization
  dynamics via magnetoelastic coupling},}\ }\href@noop {} {\bibfield  {journal}
  {\bibinfo  {journal} {Phys. Rev. B}\ }\textbf {\bibinfo {volume} {90}},\
  \bibinfo {pages} {140405(R)} (\bibinfo {year} {2014})}\BibitemShut {NoStop}%
\bibitem [{\citenamefont {Kats}\ \emph {et~al.}(2016)\citenamefont {Kats},
  \citenamefont {Linnik}, \citenamefont {Salasyuk}, \citenamefont {Rushforth},
  \citenamefont {Wang}, \citenamefont {Wadley}, \citenamefont {Akimov},
  \citenamefont {Cavill}, \citenamefont {Holy}, \citenamefont {Kalashnikova},\
  and\ \citenamefont {Scherbakov}}]{Kats_2016}%
  \BibitemOpen
  \bibfield  {author} {\bibinfo {author} {\bibfnamefont {V.}~\bibnamefont
  {Kats}}, \bibinfo {author} {\bibfnamefont {T.}~\bibnamefont {Linnik}},
  \bibinfo {author} {\bibfnamefont {A.}~\bibnamefont {Salasyuk}}, \bibinfo
  {author} {\bibfnamefont {A.}~\bibnamefont {Rushforth}}, \bibinfo {author}
  {\bibfnamefont {M.}~\bibnamefont {Wang}}, \bibinfo {author} {\bibfnamefont
  {P.}~\bibnamefont {Wadley}}, \bibinfo {author} {\bibfnamefont
  {A.}~\bibnamefont {Akimov}}, \bibinfo {author} {\bibfnamefont
  {S.}~\bibnamefont {Cavill}}, \bibinfo {author} {\bibfnamefont
  {V.}~\bibnamefont {Holy}}, \bibinfo {author} {\bibfnamefont {A.}~\bibnamefont
  {Kalashnikova}}, \ and\ \bibinfo {author} {\bibfnamefont {A.}~\bibnamefont
  {Scherbakov}},\ }\bibfield  {title} {\enquote {\bibinfo {title} {Ultrafast
  changes of magnetic anisotropy driven by laser-generated coherent and
  noncoherent phonons in metallic films},}\ }\href@noop {} {\bibfield
  {journal} {\bibinfo  {journal} {Phys. Rev. B}\ }\textbf {\bibinfo {volume}
  {93}},\ \bibinfo {pages} {214422} (\bibinfo {year} {2016})}\BibitemShut
  {NoStop}%
\bibitem [{\citenamefont {Berk}\ \emph {et~al.}(2017)\citenamefont {Berk},
  \citenamefont {Yahagi}, \citenamefont {Dhuey}, \citenamefont {Cabrini},\ and\
  \citenamefont {Schmidt}}]{Berk_2017}%
  \BibitemOpen
  \bibfield  {author} {\bibinfo {author} {\bibfnamefont {C.}~\bibnamefont
  {Berk}}, \bibinfo {author} {\bibfnamefont {Y.}~\bibnamefont {Yahagi}},
  \bibinfo {author} {\bibfnamefont {S.}~\bibnamefont {Dhuey}}, \bibinfo
  {author} {\bibfnamefont {S.}~\bibnamefont {Cabrini}}, \ and\ \bibinfo
  {author} {\bibfnamefont {H.}~\bibnamefont {Schmidt}},\ }\bibfield  {title}
  {\enquote {\bibinfo {title} {Controlling the influence of elastic eigenmodes
  on nanomagnet dynamics through pattern geometry},}\ }\href@noop {} {\bibfield
   {journal} {\bibinfo  {journal} {J. Magn. Magn. Mater}\ }\textbf {\bibinfo
  {volume} {426}},\ \bibinfo {pages} {239} (\bibinfo {year}
  {2017})}\BibitemShut {NoStop}%
\bibitem [{\citenamefont {Yang}\ \emph {et~al.}(2018)\citenamefont {Yang},
  \citenamefont {Garcia-Sanchez}, \citenamefont {Hu}, \citenamefont {Sievers},
  \citenamefont {B\"ohnert}, \citenamefont {Costa}, \citenamefont
  {Tarequzzaman}, \citenamefont {Ferreira}, \citenamefont {Bieler}, ,\ and\
  \citenamefont {Schumacher}}]{Yang_2018}%
  \BibitemOpen
  \bibfield  {author} {\bibinfo {author} {\bibfnamefont {H.}~\bibnamefont
  {Yang}}, \bibinfo {author} {\bibfnamefont {F.}~\bibnamefont
  {Garcia-Sanchez}}, \bibinfo {author} {\bibfnamefont {X.}~\bibnamefont {Hu}},
  \bibinfo {author} {\bibfnamefont {S.}~\bibnamefont {Sievers}}, \bibinfo
  {author} {\bibfnamefont {T.}~\bibnamefont {B\"ohnert}}, \bibinfo {author}
  {\bibfnamefont {J.~D.}\ \bibnamefont {Costa}}, \bibinfo {author}
  {\bibfnamefont {M.}~\bibnamefont {Tarequzzaman}}, \bibinfo {author}
  {\bibfnamefont {R.}~\bibnamefont {Ferreira}}, \bibinfo {author}
  {\bibfnamefont {M.}~\bibnamefont {Bieler}}, , \ and\ \bibinfo {author}
  {\bibfnamefont {H.~W.}\ \bibnamefont {Schumacher}},\ }\bibfield  {title}
  {\enquote {\bibinfo {title} {Excitation and coherent control of magnetization
  dynamics in magnetic tunnel junctions using acoustic pulses},}\ }\href@noop
  {} {\bibfield  {journal} {\bibinfo  {journal} {Appl. Phys. Lett}\ }\textbf
  {\bibinfo {volume} {113}},\ \bibinfo {pages} {072403} (\bibinfo {year}
  {2018})}\BibitemShut {NoStop}%
\bibitem [{\citenamefont {Deb}\ \emph {et~al.}(2018)\citenamefont {Deb},
  \citenamefont {E.~Popova}, \citenamefont {Keller}, \citenamefont {Mangin},\
  and\ \citenamefont {Malinowski}}]{Deb_2018}%
  \BibitemOpen
  \bibfield  {author} {\bibinfo {author} {\bibfnamefont {M.}~\bibnamefont
  {Deb}}, \bibinfo {author} {\bibfnamefont {M.~H.}\ \bibnamefont {E.~Popova}},
  \bibinfo {author} {\bibfnamefont {N.}~\bibnamefont {Keller}}, \bibinfo
  {author} {\bibfnamefont {S.}~\bibnamefont {Mangin}}, \ and\ \bibinfo {author}
  {\bibfnamefont {G.}~\bibnamefont {Malinowski}},\ }\bibfield  {title}
  {\enquote {\bibinfo {title} {Picosecond acoustic-excitation-driven ultrafast
  magnetization dynamics in dielectric bi-substituted yttrium iron garnet},}\
  }\href@noop {} {\bibfield  {journal} {\bibinfo  {journal} {Phys. Rev. B}\
  }\textbf {\bibinfo {volume} {98}},\ \bibinfo {pages} {174407} (\bibinfo
  {year} {2018})}\BibitemShut {NoStop}%
\bibitem [{\citenamefont {Mondal}\ \emph {et~al.}(2018)\citenamefont {Mondal},
  \citenamefont {Abeed}, \citenamefont {Dutta}, \citenamefont {De},
  \citenamefont {Sahoo}, \citenamefont {Barman},\ and\ \citenamefont
  {Bandyopadhyay}}]{Mondal_2018}%
  \BibitemOpen
  \bibfield  {author} {\bibinfo {author} {\bibfnamefont {S.}~\bibnamefont
  {Mondal}}, \bibinfo {author} {\bibfnamefont {M.}~\bibnamefont {Abeed}},
  \bibinfo {author} {\bibfnamefont {K.}~\bibnamefont {Dutta}}, \bibinfo
  {author} {\bibfnamefont {A.}~\bibnamefont {De}}, \bibinfo {author}
  {\bibfnamefont {S.}~\bibnamefont {Sahoo}}, \bibinfo {author} {\bibfnamefont
  {A.}~\bibnamefont {Barman}}, \ and\ \bibinfo {author} {\bibfnamefont
  {S.}~\bibnamefont {Bandyopadhyay}},\ }\bibfield  {title} {\enquote {\bibinfo
  {title} {Hybrid magnetodynamical modes in a single magnetostrictive
  nanomagnet on a piezoelectric substrate arising from magnetoelastic
  modulation of precessional dynamics},}\ }\href@noop {} {\bibfield  {journal}
  {\bibinfo  {journal} {ACS Appl. Mater. Interfaces}\ }\textbf {\bibinfo
  {volume} {10}},\ \bibinfo {pages} {43970} (\bibinfo {year}
  {2018})}\BibitemShut {NoStop}%
\bibitem [{\citenamefont {Hashimoto}\ \emph {et~al.}(2018)\citenamefont
  {Hashimoto}, \citenamefont {Bossini}, \citenamefont {Johansen}, \citenamefont
  {Saitoh}, \citenamefont {Kirilyuk},\ and\ \citenamefont
  {Rasing}}]{Hashimoto_2018}%
  \BibitemOpen
  \bibfield  {author} {\bibinfo {author} {\bibfnamefont {Y.}~\bibnamefont
  {Hashimoto}}, \bibinfo {author} {\bibfnamefont {D.}~\bibnamefont {Bossini}},
  \bibinfo {author} {\bibfnamefont {T.~H.}\ \bibnamefont {Johansen}}, \bibinfo
  {author} {\bibfnamefont {E.}~\bibnamefont {Saitoh}}, \bibinfo {author}
  {\bibfnamefont {A.}~\bibnamefont {Kirilyuk}}, \ and\ \bibinfo {author}
  {\bibfnamefont {T.}~\bibnamefont {Rasing}},\ }\bibfield  {title} {\enquote
  {\bibinfo {title} {Frequency and wavenumber selective excitation of spin
  waves through coherent energy transfer from elastic waves},}\ }\href@noop {}
  {\bibfield  {journal} {\bibinfo  {journal} {Phys. Rev. B}\ }\textbf {\bibinfo
  {volume} {97}},\ \bibinfo {pages} {140404} (\bibinfo {year}
  {2018})}\BibitemShut {NoStop}%
\bibitem [{\citenamefont {Limonov}\ \emph {et~al.}(2017)\citenamefont
  {Limonov}, \citenamefont {Rybin}, \citenamefont {Poddubny},\ and\
  \citenamefont {Kivshar}}]{Limonov_2017}%
  \BibitemOpen
  \bibfield  {author} {\bibinfo {author} {\bibfnamefont {M.~F.}\ \bibnamefont
  {Limonov}}, \bibinfo {author} {\bibfnamefont {M.~V.}\ \bibnamefont {Rybin}},
  \bibinfo {author} {\bibfnamefont {A.~N.}\ \bibnamefont {Poddubny}}, \ and\
  \bibinfo {author} {\bibfnamefont {Y.~S.}\ \bibnamefont {Kivshar}},\
  }\bibfield  {title} {\enquote {\bibinfo {title} {Fano resonances in
  photonics},}\ }\href@noop {} {\bibfield  {journal} {\bibinfo  {journal}
  {Nature Photon}\ }\textbf {\bibinfo {volume} {11}},\ \bibinfo {pages} {543}
  (\bibinfo {year} {2017})}\BibitemShut {NoStop}%
\bibitem [{\citenamefont {Comstock}\ and\ \citenamefont
  {Auld}(1963)}]{Comstock_1963}%
  \BibitemOpen
  \bibfield  {author} {\bibinfo {author} {\bibfnamefont {R.~L.}\ \bibnamefont
  {Comstock}}\ and\ \bibinfo {author} {\bibfnamefont {B.~A.}\ \bibnamefont
  {Auld}},\ }\bibfield  {title} {\enquote {\bibinfo {title} {Parametric
  coupling of the magnetization and strain in a ferrimagnet. i. parametric
  excitation of magnetostatic and elastic modes},}\ }\href@noop {} {\bibfield
  {journal} {\bibinfo  {journal} {J. Appl. Phys.}\ }\textbf {\bibinfo {volume}
  {34}},\ \bibinfo {pages} {1461} (\bibinfo {year} {1963})}\BibitemShut
  {NoStop}%
\bibitem [{\citenamefont {Kamra}\ \emph {et~al.}(2015)\citenamefont {Kamra},
  \citenamefont {Keshtgar}, \citenamefont {Yan},\ and\ \citenamefont
  {Bauer}}]{Kamra_2015}%
  \BibitemOpen
  \bibfield  {author} {\bibinfo {author} {\bibfnamefont {A.}~\bibnamefont
  {Kamra}}, \bibinfo {author} {\bibfnamefont {H.}~\bibnamefont {Keshtgar}},
  \bibinfo {author} {\bibfnamefont {P.}~\bibnamefont {Yan}}, \ and\ \bibinfo
  {author} {\bibfnamefont {G.~E.~W.}\ \bibnamefont {Bauer}},\ }\bibfield
  {title} {\enquote {\bibinfo {title} {Coherent elastic excitation of spin
  waves},}\ }\href@noop {} {\bibfield  {journal} {\bibinfo  {journal} {Phys.
  Rev. B}\ }\textbf {\bibinfo {volume} {91}},\ \bibinfo {pages} {104409}
  (\bibinfo {year} {2015})}\BibitemShut {NoStop}%
\bibitem [{\citenamefont {Callen}\ and\ \citenamefont
  {Callen}(1965)}]{Callen_1965}%
  \BibitemOpen
  \bibfield  {author} {\bibinfo {author} {\bibfnamefont {E.}~\bibnamefont
  {Callen}}\ and\ \bibinfo {author} {\bibfnamefont {H.~B.}\ \bibnamefont
  {Callen}},\ }\bibfield  {title} {\enquote {\bibinfo {title}
  {Magnetostriction, forced magnetostriction, and anomalous thermal expansion
  in ferromagnets},}\ }\href@noop {} {\bibfield  {journal} {\bibinfo  {journal}
  {Phys. Rev}\ }\textbf {\bibinfo {volume} {139}},\ \bibinfo {pages} {A455}
  (\bibinfo {year} {1965})}\BibitemShut {NoStop}%
\bibitem [{\citenamefont {Brekhovskikh}\ and\ \citenamefont
  {Godin}(1997)}]{Brekhovskikh_1997}%
  \BibitemOpen
  \bibfield  {author} {\bibinfo {author} {\bibfnamefont {L.~M.}\ \bibnamefont
  {Brekhovskikh}}\ and\ \bibinfo {author} {\bibfnamefont {O.~A.}\ \bibnamefont
  {Godin}},\ }\href@noop {} {\emph {\bibinfo {title} {Acoustics of Layered
  Media}}}\ (\bibinfo  {publisher} {Berlin, Heidelberg: Springer},\ \bibinfo
  {year} {1997})\BibitemShut {NoStop}%
\bibitem [{\citenamefont {Born}\ and\ \citenamefont {Wolf}(1964)}]{Born_1964}%
  \BibitemOpen
  \bibfield  {author} {\bibinfo {author} {\bibfnamefont {M.}~\bibnamefont
  {Born}}\ and\ \bibinfo {author} {\bibfnamefont {E.}~\bibnamefont {Wolf}},\
  }\href@noop {} {\emph {\bibinfo {title} {Principles of Optics}}}\ (\bibinfo
  {publisher} {Oxford, New York: Pergamon},\ \bibinfo {year}
  {1964})\BibitemShut {NoStop}%
\bibitem [{\citenamefont {Landau}\ and\ \citenamefont
  {Lifshitz}(1965)}]{Landau_1965}%
  \BibitemOpen
  \bibfield  {author} {\bibinfo {author} {\bibfnamefont {L.}~\bibnamefont
  {Landau}}\ and\ \bibinfo {author} {\bibfnamefont {E.~M.}\ \bibnamefont
  {Lifshitz}},\ }\href@noop {} {\emph {\bibinfo {title} {Quantum Mechanics}}}\
  (\bibinfo  {publisher} {Oxford, New York: Pergamon},\ \bibinfo {year}
  {1965})\ Chap.\ \bibinfo {chapter} {145}\BibitemShut {NoStop}%
\bibitem [{\citenamefont {Klaiman}(2017)}]{Klaiman_2017}%
  \BibitemOpen
  \bibfield  {author} {\bibinfo {author} {\bibfnamefont {S.}~\bibnamefont
  {Klaiman}},\ }\bibfield  {title} {\enquote {\bibinfo {title} {Reflections on
  one dimensionl transmission},}\ }\href@noop {} {\bibfield  {journal}
  {\bibinfo  {journal} {Chem. Phys.}\ }\textbf {\bibinfo {volume} {482}},\
  \bibinfo {pages} {277} (\bibinfo {year} {2017})}\BibitemShut {NoStop}%
\bibitem [{\citenamefont {Emori}\ \emph {et~al.}(2017)\citenamefont {Emori},
  \citenamefont {Gray}, \citenamefont {Jeon}, \citenamefont {Peoples},
  \citenamefont {Schmitt}, \citenamefont {Mahalingam}, \citenamefont {Hill},
  \citenamefont {McConney}, \citenamefont {Gray}, \citenamefont {Alaan},
  \citenamefont {Bornstein}, \citenamefont {Shafer}, \citenamefont {N'Diaye},
  \citenamefont {Arenholz}, \citenamefont {Haugstad}, \citenamefont {Meng},
  \citenamefont {Yang}, \citenamefont {Li}, \citenamefont {Mahat},
  \citenamefont {Cahill}, \citenamefont {Dhagat}, \citenamefont {Jander},
  \citenamefont {Sun}, \citenamefont {Suzuki},\ and\ \citenamefont
  {Howe}}]{Emori_2017}%
  \BibitemOpen
  \bibfield  {author} {\bibinfo {author} {\bibfnamefont {S.}~\bibnamefont
  {Emori}}, \bibinfo {author} {\bibfnamefont {B.}~\bibnamefont {Gray}},
  \bibinfo {author} {\bibfnamefont {H.-M.}\ \bibnamefont {Jeon}}, \bibinfo
  {author} {\bibfnamefont {J.}~\bibnamefont {Peoples}}, \bibinfo {author}
  {\bibfnamefont {M.}~\bibnamefont {Schmitt}}, \bibinfo {author} {\bibfnamefont
  {K.}~\bibnamefont {Mahalingam}}, \bibinfo {author} {\bibfnamefont
  {M.}~\bibnamefont {Hill}}, \bibinfo {author} {\bibfnamefont {M.}~\bibnamefont
  {McConney}}, \bibinfo {author} {\bibfnamefont {M.}~\bibnamefont {Gray}},
  \bibinfo {author} {\bibfnamefont {U.}~\bibnamefont {Alaan}}, \bibinfo
  {author} {\bibfnamefont {A.}~\bibnamefont {Bornstein}}, \bibinfo {author}
  {\bibfnamefont {P.}~\bibnamefont {Shafer}}, \bibinfo {author} {\bibfnamefont
  {A.}~\bibnamefont {N'Diaye}}, \bibinfo {author} {\bibfnamefont
  {E.}~\bibnamefont {Arenholz}}, \bibinfo {author} {\bibfnamefont
  {G.}~\bibnamefont {Haugstad}}, \bibinfo {author} {\bibfnamefont {K.-Y.}\
  \bibnamefont {Meng}}, \bibinfo {author} {\bibfnamefont {F.}~\bibnamefont
  {Yang}}, \bibinfo {author} {\bibfnamefont {D.}~\bibnamefont {Li}}, \bibinfo
  {author} {\bibfnamefont {S.}~\bibnamefont {Mahat}}, \bibinfo {author}
  {\bibfnamefont {D.}~\bibnamefont {Cahill}}, \bibinfo {author} {\bibfnamefont
  {P.}~\bibnamefont {Dhagat}}, \bibinfo {author} {\bibfnamefont
  {A.}~\bibnamefont {Jander}}, \bibinfo {author} {\bibfnamefont
  {N.}~\bibnamefont {Sun}}, \bibinfo {author} {\bibfnamefont {Y.}~\bibnamefont
  {Suzuki}}, \ and\ \bibinfo {author} {\bibfnamefont {B.}~\bibnamefont
  {Howe}},\ }\bibfield  {title} {\enquote {\bibinfo {title} {Coexistence of low
  damping and strong magnetoelastic coupling in epitaxial spinel ferrite thin
  films},}\ }\href@noop {} {\bibfield  {journal} {\bibinfo  {journal} {Adv.
  Mater.}\ }\textbf {\bibinfo {volume} {29}},\ \bibinfo {pages} {1701130}
  (\bibinfo {year} {2017})}\BibitemShut {NoStop}%
\bibitem [{\citenamefont {Kim}(2012)}]{Kim_2012}%
  \BibitemOpen
  \bibfield  {author} {\bibinfo {author} {\bibfnamefont {J.~V.}\ \bibnamefont
  {Kim}},\ }\href@noop {} {\emph {\bibinfo {title} {Spin Torque
  Oscillators}}},\ Vol.~\bibinfo {volume} {63}\ (\bibinfo  {publisher}
  {Academic Press},\ \bibinfo {year} {2012})\ Chap.~\bibinfo {chapter}
  {4}\BibitemShut {NoStop}%
\end{thebibliography}%

\end{document}